\documentclass[pre,preprint,showpacs]{revtex4}

\usepackage[latin1]{inputenc}
\usepackage{amssymb,amsmath}
\usepackage{graphicx}

\begin{document}
\title{Shear viscosity for a moderately dense granular binary mixture}
\author{Vicente Garz\'{o}}
\email[E-mail: ]{vicenteg@unex.es}
\address{Departamento de F\'{\i}sica, Universidad de Extremadura, E-06071
Badajoz, Spain}
\author{Jos\'e Mar\'{\i}a Montanero}
\email[E-mail: ]{jmm@unex.es}
\address{Departamento de Electr\'onica e Ingenier\'{\i}a Electromec\'anica,
Universidad de Extremadura, E-06071 Badajoz, Spain}

\begin{abstract} 
The shear viscosity for a moderately dense granular binary mixture of smooth hard spheres undergoing uniform shear flow is determined. The basis for the analysis is the Enskog kinetic equation, solved first analytically by the Chapman-Enskog method up to first order in the shear rate for unforced systems as well as for systems driven by a Gaussian thermostat. As in the elastic case, practical evaluation requires a Sonine polynomial approximation. In the leading order, we determine the shear viscosity in terms of the control parameters of the problem: solid fraction, composition, mass ratio, size ratio and restitution coefficients. Both kinetic and collisional transfer contributions to the shear viscosity are considered. To probe the accuracy of the Chapman-Enskog results, the Enskog equation is then numerically solved for systems driven by a Gaussian thermostat by means of an extension to dense gases of the well-known Direct Simulation Monte Carlo (DSMC) method for dilute gases. The comparison between theory and simulation shows in general an excellent agreement over a wide region of the parameter space. 
\end{abstract} 

\draft
\date{\today}
\maketitle

\section{Introduction}
\label{sec1}

An usual way of capturing the dissipative nature of granular media is through an idealized fluid of smooth, inelastic hard spheres. Despite the simplicity of the model, it has been shown to be quite useful in describing the dynamics of granular materials under rapid flow conditions\cite{C90,PL01}. The essential difference from molecular fluids is the absence of energy conservation, leading to both obvious and subtle modifications of the Navier-Stokes hydrodynamic equations. Although many efforts have been made in the past few years in the understanding of granular fluids, the derivation of the form of the transport coefficients remains a topic of interest and controversy.  This problem has been addressed using the inelastic Boltzmann equation or its dense fluid generalization, the Enskog equation. Assuming the existence of a normal solution for sufficiently long space and time scales, the Chapman-Enskog method \cite{CC70}, conveniently adapted to inelastic collisions, has been applied to get the Navier-Stokes transport coefficients. For a monocomponent gas at low-density, the above  coefficients have been explicitly determined as functions of the restitution coefficient \cite{BDKS98,BC01,GM02} from  approximate solutions of the corresponding kinetic equations. The accuracy of these approximate results has been then confirmed by computer simulations\cite{GM02,BRC99}. The analysis for dilute gases has been also extended to finite  densities in the context of the revised Enskog kinetic theory (RET) \cite{GD99a}. This hydrodynamic theory succesfully models the density and temperature profiles obtained in a recent experimental study of a three-dimensional system of mustard seeds fluidized by vertical container vibrations\cite{mustard}. 

The majority of the studies on granular fluids are confined to monocomponent systems, where the particles are of the same mass and size. However, a real granular system is always characterized by some degrees of polydispersity in density and size, which often leads to segregation of an otherwise homogeneous mixture.  Needless to say, the analysis of transport for multicomponent systems is much more involved than for a monocomponent gas. Not only the number of transport coefficients is higher but also they are functions of parameters such as the mole fractions, the mass ratios, the size ratios and the restitution coefficients. For this reason,  most of the previous studies \cite{Jenkins} are restricted to {\em nearly} elastic spheres. In addition, they usually assume energy equipartition so that the partial temperatures $T_i$ are made equal to the global granular temperature $T$. Nevertheless, recent experiments of vibrated mixtures in three \cite{WP02} and two \cite{FM02} dimensions clearly show the breakdown of energy equipartition. Related findings have also been reported by using kinetic theory tools \cite{GD99b,BT02} and computer simulations\cite{MG02,DHGD02}. To the best of our knowledge, the only kinetic theory derivation of hydrodynamics for a granular binary mixture at low-density which takes into account nonequipartition of granular energy has been made by Garz\'o and Dufty\cite{GD02}. They solved the Boltzmann equation by applying the Chapman-Enskog method to obtain the Navier-Stokes equations and detailed expressions for the transport coefficients. In the case of the shear viscosity, the reliability of the kinetic theory predictions have also been assessed \cite{MG03} in a wide parameter space by comparing those predictions with the results obtained from  a numerical solution of the Boltzmann equation by means of the Direct Simulation Monte Carlo (DSMC) method\cite{B94}. The comparison shows an excellent agreement between theory and simulation. 

The objective here is to extend the analysis carried out in Ref.\ \cite{MG03} for the shear viscosity to higher densities by using the RET. The RET for elastic collisions \cite{BE73} is known to be an accurate theory over the entire fluid domain. Its generalization to inelastic collisions is straightforward (see, for example, Ref.\ \cite{BDS97}) and the Chapman-Enskog method can be applied to obtain the transport coefficients. However, the derivation of the hydrodynamic equations for a binary mixture described by the RET is more complicated than in the case of the Boltzmann equation, due mainly to the technical difficulties associated with the spatial dependence of the pair correlation function. To simplify this analysis, here attention is restricted to the special hydrodynamic state of uniform shear flow (USF). At a macroscopic level, this state is characterized by constant partial densities $n_i$, uniform temperature $T$ and a linear flow velocity profile ${\bf u}_i=ay{\bf {\widehat{x}}}$, $a$ being the constant shear rate. For this particular problem the RET reduces to the original phenomonological kinetic theory proposed by Enskog\cite{FK72}. We solve the Enskog equation up to first order in the shear rate and evaluate both kinetic and collisional transfer contributions to the shear viscosity. This transport coefficient is expressed  in terms of the solution of a set of coupled linear integral equations, which are then solved approximately (first Sonine polynomial approximation) just as in the case of elastic collisions. As done in the low-density analysis\cite{MG03}, the Sonine solution is compared with a numerical solution of the RET by using the Enskog Simulation Monte Carlo (ESMC) method\cite{MS96}, which is an extension to the Enskog equation of the well-known DSMC method\cite{B94}. 

In a molecular fluid under USF, unless a thermostating force is introduced, the temperature grows in time due to viscous heating.  As a consequence, the average collision frequency $\nu(t)\propto T^{1/2}(t)$ increases with time and the reduced shear rate $a^*=a/\nu$ goes to zero  in the long time limit. This fact allows one to identify in the simulation the Navier-Stokes shear viscosity coefficient $\eta$ for sufficiently long times. This route has been shown to be quite efficient to measure $\eta$  for dilute and dense gases \cite{MS96,NO79}.  For a granular fluid, there is an additional energy sink term in the balance equation for the temperature competing with the viscous heating term. However, if the effect of the former term is exactly compensated by for the action of an external driving force, the viscous heating prevails and the shear viscosity can be again identified in the limit $a^* \to 0$, just as in the elastic case. This was the procedure followed in Ref.\ \cite{MG03} to measure $\eta$ from the simulation in the long time limit. It must be noted that the value of $\eta$ calculated in this way (driven case) not necessarily coincides with the value of the shear viscosity obtained in the free cooling case (unforced case). 

There are several motivations for this study. First,  we want to assess to what extent the previous results obtained for the low-density regime are indicative of what happens for finite densities. Second, the comparison between theory and simulation allows one to check the degree of reliability of the approximate Sonine solution over a wide region of parameter space. Finally, by extending the Boltzmann analysis to higher densities, comparison with molecular dynamics simulations become practical. This comparison would determine the validity (or limitations) of the kinetic and hydrodynamic descriptions for granular flow. Such a test is essential to clarify the frequently made speculation that the above descriptions of granular flow are limited to weak dissipation. Some previous comparisons\cite{DHGD02,LBD02} support the hydrodynamic description, beyond complications due to possible instabilities.

The plan of the paper is as follows. In Sec.\ \ref{sec2} we review  the Enskog theory and deduce the associated macroscopic conservation equations. The Chapman-Enskog method is applied in Sec.\ \ref{sec3} to solve the Enskog equation in the USF state through first order in the shear rate.  An explicit expression for the shear viscosity coefficient is obtained in  Sec.\ \ref{sec3bis} by using a lowest order expansion in Sonine polynomials. This transport coefficient is given in terms of the restitution coefficients, the temperature, the solid fraction, and the parameters characterizing the mixture (masses, sizes, concentrations). Section \ref{sec4} deals with the Monte Carlo simulation of the Enskog equation particularized to USF. The comparison between theory and simulation is carried out in Sec.\ \ref{sec5}, while a brief discussion on the relevance of the results obtained is given in Sec.\ \ref{sec6}.

\section{Enskog kinetic theory and conservation laws}
\label{sec2}

We consider a binary mixture of smooth hard spheres of masses $m_{1}$ and $m_{2} $, and diameters $\sigma _{1}$ and $\sigma _{2}$. The inelasticity of collisions among all pairs is characterized by three independent constant coefficients of normal restitution $\alpha_{11}$, $\alpha_{22}$, and $ \alpha_{12}=\alpha_{21}$, where $\alpha_{ij}$ is the restitution coefficient for collisions between particles of species $i$ and $j$. Due to the intrinsic dissipative character of collisions, in order to keep the system under rapid flow conditions it is usual to introduce an external driving force (thermostat) which does work to compensate for the collisional loss of energy. This mechanism of energy input (different from those of shear flows or flows through vertical pipes) has been used for many authors in the past years to study different problems, such as non-Gaussian properties of the velocity distribution function \cite{NE98,MS00}, long-range correlations\cite{NETP99}, collisional statistics and short-scale structure\cite{PTNE02}, or transport properties\cite{BSSS99}. In this paper, for simplicity,  we introduce a deterministic force proportional to the peculiar velocity ${\bf V}$ (Gaussian thermostat). This thermostat has been frequently employed in nonequilibrium molecular dynamics simulations of elastic particles\cite{EM90}.  Under these conditions,  the Enskog kinetic equation for the one-particle velocity distribution function of species $i$ is given by 
\begin{equation} 
\left( \partial _{t}+{\bf v}_{1}\cdot \nabla \right) 
f_{i}+\frac{1}{2}\xi
\frac{\partial}{\partial {\bf v}_1}\cdot \left({\bf V}_1f_i\right)
=\sum_{j=1}^2J_{ij}^{\text{E}}\left[ {\bf r}, {\bf v}_{1}|f_{i}(t),f_{j}(t)\right] \;,
\label{2.1} 
\end{equation} 
where the constant $\xi$ is chosen to be the same for both species. Here, ${\bf V}_1={\bf v}_1-{\bf u}$, ${\bf u}$ being the flow velocity. The Enskog collision operator $J_{ij}^{\text{E}}[f_i,f_j]$ is\cite{BDS97} 
\begin{eqnarray} 
\label{2.2}
J_{ij}^{\text{E}}\left[ {\bf r}, {\bf v}_{1}|f_{i},f_{j}\right] &=&\sigma _{ij}^{2}\int d{\bf v} 
_{2}\int d\widehat{\boldsymbol{\sigma}}\,\Theta (\widehat{{\boldsymbol {\sigma }}} 
\cdot {\bf g})(\widehat{\boldsymbol {\sigma }}\cdot {\bf g})  \nonumber 
\\ &&\times \left[ \alpha _{ij}^{-2}\chi_{ij}({\bf r},{\bf r}-\boldsymbol {\sigma }_{ij}) 
f_i({\bf r}, {\bf v}_1';t)f_j({\bf r}-\boldsymbol {\sigma}_{ij}, {\bf v}_2';t)\right.
\nonumber\\
& & \left.
-\chi_{ij}({\bf r},{\bf r}+\boldsymbol {\sigma }_{ij}) 
f_i({\bf r}, {\bf v}_1;t)f_j({\bf r}+\boldsymbol {\sigma }_{ij}, {\bf v}_2;t)\right],
\end{eqnarray} 
where $\boldsymbol {\sigma}_{ij}=\sigma_{ij} \widehat{\boldsymbol {\sigma }}$, with $\sigma _{ij}=\left( \sigma _{i}+\sigma _{j}\right) /2$ and $\widehat{\boldsymbol {\sigma}}$ is a unit vector directed along the line of centers from the sphere of species $i$ to the sphere of species $j$ upon collision (i.e. at contact). In addition,  $\Theta $ is 
the Heaviside step function, and ${\bf g}={\bf v}_{1}-{\bf v}_{2}$. The 
primes on the velocities denote the initial values $\{{\bf v}_{1}^{\prime},  
{\bf v}_{2}^{\prime }\}$ that lead to $\{{\bf v}_{1},{\bf v}_{2}\}$ 
following a binary collision:  
\begin{equation} 
{\bf v}_{1}^{\prime }={\bf v}_{1}-\mu _{ji}\left( 1+\alpha _{ij}^{-1}\right) 
(\widehat{{\boldsymbol {\sigma }}}\cdot {\bf g})\widehat{{\boldsymbol {\sigma }}} 
,\quad {\bf v}_{2}^{\prime }={\bf v}_{2}+\mu _{ij}\left( 1+\alpha 
_{ij}^{-1}\right) (\widehat{{\boldsymbol {\sigma }}}\cdot {\bf g})\widehat{ 
\boldsymbol {\sigma}} , \label{2.3} 
\end{equation} 
where $\mu _{ij}=m_{i}/\left( m_{i}+m_{j}\right)$.  Finally, $\chi_{ij}[{\bf r},{\bf r}+\boldsymbol {\sigma}_{ij}|\{n_\ell\}] $ is the equilibrium pair correlation function of two hard spheres, one of species $i$ and the other of species $j$, at contact, i.e., when the distance between their centers is $\sigma_{ij}$. In the original phenomenological kinetic theory of Enskog\cite{FK72} (which is usually referred to as the standard Enskog theory), the $\chi_{ij}$ are the same {\em functions} of the densities $\{n_\ell\}$ as in a fluid mixture in {\em uniform} equilibrium. Here, 
\begin{equation} 
n_{i}=\int d{\bf v}f_{i}({\bf v}) \label{2.4} 
\end{equation} 
is the number density of species $i$. On the other hand, this choice for $\chi_{ij}$ leads to some inconsistencies with irreversible thermodynamics.  In order to resolve it, van Beijeren and Ernst\cite{BE73} proposed  an alternative generalization to the Enskog equation for mixtures, which is usually referred to as the revised Enskog theory (RET).  In the RET, the $\chi_{ij}$ are the same {\em functionals} of the densities $\{n_\ell\}$ as in a fluid in {\em nonuniform} equilibrium. This fact increases considerably the technical difficulties involved in the derivation of the general hydrodynamic equations from the RET\cite{MCK83,MG93}, unless the partial densities are uniform. 

The macroscopic balance equations for the particle number of each species, the total momentum and the total energy follow directly from Eq.\ (\ref{2.1}) by multiplying by 1, $m_i {\bf v}$, and $\case{1}{2}m_iv^2$, respectively, integrating over ${\bf v}$, and summing over $i$. They are given by \begin{equation} 
\frac{\partial}{\partial t}n_{i}+\nabla \cdot(n_i {\bf u})+\frac{\nabla \cdot {\bf j}_{i}}{m_{i}} 
=0\;,  \label{2.7} 
\end{equation} 
\begin{equation} 
\frac{\partial}{\partial t}{\bf u}+{\bf u}\cdot \nabla {\bf u}+\rho ^{-1}\nabla \cdot {\sf P}=0\;,  \label{2.8} 
\end{equation} 
\begin{equation} 
\frac{\partial}{\partial t}T+{\bf u}\cdot \nabla T -\frac{T}{n}\sum_{i=1}^2\frac{\nabla \cdot {\bf j}_{i}}{m_{i}}+\frac{2}{3n} 
\left( \nabla \cdot {\bf q}+{\sf P}:\nabla {\bf u}\right) 
=-(\zeta-\xi)T\;. \label{2.9} 
\end{equation} 
Here, $\zeta$ is the cooling rate due to inelastic collisions among all species. The flow velocity ${\bf u}$ and the ``granular'' temperature $T$ are defined by 
\begin{equation} 
\rho {\bf u}=\sum_{i=1}^2\int d {\bf v}m_{i}{\bf v}f_{i}({\bf v})\;,  \label{2.10} 
\end{equation} 
\begin{equation} 
nT=\sum_{i=1}^2\int d{\bf v}\frac{m_{i}}{3}V^{2}f_{i}({\bf v})\;, 
\label{2.11} 
\end{equation} 
where $n=n_{1}+n_{2}$ is the total number density, and $\rho 
=m_{1}n_{1}+m_{2}n_{2}$ is the total mass density. The mass flux ${\bf j}_i$ for species $i$ relative to the local flow is given by 
\begin{equation} 
{\bf j}_{i}=m_{i}\int d{\bf v}\,{\bf V}\,f_{i}({\bf v}).
\label{2.13} 
\end{equation} 
The pressure tensor ${\sf P}$ and the heat flux ${\bf q}$ have both {\em kinetic} and {\em collisional transfer contributions}, i.e., ${\sf P}={\sf P}^{\text{k}}+{\sf P}^{\text{c}}$ and  ${\bf q}={\bf q}^{\text{k}}+{\bf q}^{\text{c}}$. The kinetic contributions are given by  
\begin{equation} 
{\sf P}^{\text{k}}=\sum_{i=1}^2\,\int d{\bf v}\,m_{i}{\bf V}{\bf V}\,f_{i}({\bf  
v}) , \label{2.14} 
\end{equation} 
\begin{equation} 
{\bf q}^{\text{k}}=\sum_{i=1}^2\,\int d{\bf v}\,\case{1}{2}m_{i}V^{2}{\bf V} 
\,f_{i}({\bf v}), \label{2.15} 
\end{equation} 
while the collisional transfer contributions to the pressure tensor  and the heat flux are, respectively,
\begin{eqnarray}
\label{2.17}
{\sf P}^{\text{c}}({\bf r},t)&=&\sum_{i=1}^2\sum_{j=1}^2\sigma_{ij}^{3}\frac{m_im_j}{m_i+m_j}\frac{1+\alpha_{ij}}{2} 
\int d{\bf v}_1\int d{\bf v} _{2}\int d\widehat{\boldsymbol {\sigma }}\,\Theta (\widehat{{\boldsymbol {\sigma }}} 
\cdot {\bf g})(\widehat{\boldsymbol {\sigma }}\cdot {\bf g})^2 \widehat{{\boldsymbol {\sigma }}} \widehat{{\boldsymbol {\sigma }}}  \nonumber 
\\ &&\times  \int_{0}^{1} d\lambda \chi_{ij}[{\bf r}-(1-\lambda)\boldsymbol {\sigma}_{ij},{\bf r}+\lambda \mathbf{\sigma}_{ij}]f_i({\bf r}- (1-\lambda)\boldsymbol {\sigma}_{ij},{\bf v}_1;t)f_j({\bf r}+
\lambda \boldsymbol {\sigma}_{ij}, {\bf v}_2;t),\nonumber\\
\end{eqnarray}
\begin{eqnarray}
\label{2.18}
{\bf q}^{\text{c}}({\bf r},t)&=&\sum_{i=1}^2\sum_{j=1}^2\sigma_{ij}^{3}\frac{m_im_j}{m_i+m_j}
\frac{1+\alpha_{ij}}{2} \int d{\bf v}_1\int d{\bf v} 
_{2}\int d\widehat{\boldsymbol {\sigma }}\,\Theta (\widehat{{\boldsymbol {\sigma }}} 
\cdot {\bf g})(\widehat{\boldsymbol {\sigma }}\cdot {\bf g})^2 \widehat{{\boldsymbol {\sigma }}}\nonumber\\
& & \times   
\left[(\widehat{\boldsymbol {\sigma }}\cdot {\bf G}_{ij})+\frac{1}{4}\frac{m_j-m_i}{m_i+m_j}(1-\alpha_{ij})
(\widehat{\boldsymbol {\sigma }}\cdot {\bf g})\right]\nonumber\\ 
&&\times  \int_{0}^{1} d\lambda \chi_{ij}[{\bf r}-(1-\lambda)\boldsymbol {\sigma}_{ij},{\bf r}+\lambda \mathbf{\sigma}_{ij}) f_i({\bf r}- (1-\lambda)\boldsymbol {\sigma}_{ij},{\bf v}_1;t)f_j({\bf r}+\lambda \boldsymbol {\sigma}_{ij}, {\bf v}_2;t). \nonumber\\
\end{eqnarray}
Here, ${\bf G}_{ij}=\mu_{ij}{\bf V}_1+\mu_{ji}{\bf V}_2$ is the center-of-mass velocity. Finally, the cooling rate  $\zeta$ in Eq.\ (\ref{2.9}) is  
\begin{eqnarray}
\label{2.19}
\zeta({\bf r},t)&=&\frac{1}{6nT}\sum_{i=1}^2\sum_{j=1}^2
\sigma_{ij}^{2}\frac{m_im_j}{m_i+m_j}(1-\alpha_{ij}^2)
\int d{\bf v}_1\int d{\bf v} _{2}\int d\widehat{\boldsymbol {\sigma }}\,\Theta (\widehat{{\boldsymbol {\sigma }}} 
\cdot {\bf g})(\widehat{\boldsymbol {\sigma }}\cdot {\bf g})^3 \nonumber\\
& & \times
\chi_{ij}({\bf r},{\bf r}+\mathbf{\sigma}_{ij}) 
f_i({\bf r}, {\bf v}_1;t)f_j({\bf r}+\boldsymbol {\sigma}_{ij}, {\bf v}_2;t).
\end{eqnarray}
The derivation of Eqs.\ (\ref{2.17})--(\ref{2.19}) is given in  Appendix \ref{appA}. The collisional transfer contributions are due to the delocalization of the colliding pair and the additional density dependence of the RET. They vanish in the low density limit but dominate at high densities. In the case of mechanically equivalent particles ($m_1=m_2$, 
$\alpha_{11}=\alpha_{22}=\alpha_{12}\equiv\alpha$, $\sigma_1=\sigma_2
\equiv\sigma$, $\chi_{ij}\equiv\chi$), Eqs.\ (\ref{2.17})--(\ref{2.19}) reduce to those previously obtained in the monocomponent case\cite{BDS97}. 

The balance equations contain the mass flux, the heat flux, and the pressure tensor as specific averages over the distribution functions $f_i$. The Chapman-Enskog method \cite{CC70} provides a solution of the RET for states with small spatial variations in the form 
\begin{equation}
\label{2.19bis}
f_i({\bf r}, {\bf v}_1,t)=f_i[{\bf v}_1|n_1({\bf r},t), T({\bf r},t), {\bf u}({\bf r},t)].
\end{equation}  
This means that all space and time dependence of $f_i({\bf r}, {\bf v}_1,t)$ occurs entirely through a functional dependence on the hydrodynamic fields. Such a solution is called {\em normal} and it is the basis for a fluid dynamics description of granular materials. Regarding the energy input mechanism  we see that, according to the energy balance equation (\ref{2.9}), the existence of a driving with the choice $\xi=\zeta$ compensates for the cooling effect due to the inelasticity of collisions. In that case, the macroscopic balance equations look like those of a conventional mixture with elastic collisions, although the transport coefficients entering in the constitutive equations are in general different from those of a gas of elastic particles. However, the evaluation of the complete transport coefficients of the RET for a multicomponent granular mixture is a very hard task and here  we will pay attention to the shear viscosity coefficient only. Specifically, this coefficient  will be determined in a particular simple situation (uniform shear flow) where the velocity field is the only inhomogeneity present in the system. In this case, the $\chi_{ij}$ are uniform so that the standard and revised Enskog theories are equivalent in this problem. Further, the simplicity of this state allows us to check our theoretical predictions for the shear viscosity with those obtained from a numerical solution of the corresponding Enskog equation.

\section{Shear viscosity of a dense granular binary mixture}
\label{sec3}

As said above, we want to solve the Enskog equation (\ref{2.1}) in the specific state of the uniform shear flow (USF). In this state, the partial densities $n_i$ and the temperature $T$ are uniform, while the velocity field is due to a simple shear
\begin{equation}
\label{3.1}
{\bf u}_1={\bf u}_2={\bf u}=ay{\bf {\widehat{x}}}, \quad a=\frac{\partial u_x}{\partial y}=\text{constant}.
\end{equation}
The temperature changes in time due to the competition between two mechanisms: on the one hand, viscous heating and, on the other hand, energy dissipation in collisions. Under these conditions, the mass and heat fluxes vanish by symmetry reasons and the (uniform) pressure tensor ${\sf P}$ is the only nonzero flux of the problem.  The relevant balance equation is that for the temperature (\ref{2.9}), which reduces to
\begin{equation}
\label{3.3}
\partial_tT+\frac{2}{3n}aP_{xy}=-\left(\zeta-\xi\right)T. 
\end{equation}

At a microscopic level, the USF is generated by Lees-Edwards boundary conditions\cite{LE72} which are simply periodic boundary conditions in the local Lagrangian frame ${\bf V}={\bf v}-{\sf a}\cdot {\bf r}$ and ${\bf R}={\bf r}-{\sf a}\cdot {\bf r}t$. Here, ${\sf a}$ is the tensor with elements $a_{\alpha\beta}=a\delta_{\alpha x}\delta_{\beta y}$. In terms of the above variables, the velocity distribution functions are uniform\cite{DSBR86}
\begin{equation}
\label{3.3.1} 
f_i({\bf r},{\bf v},t)=f_i({\bf V},t),
\end{equation} 
and the Enskog equation takes the form
\begin{equation} 
\partial _{t}f_i-aV_y\frac{\partial}{\partial V_x} 
f_{i}+\frac{1}{2}\xi \frac{\partial}{\partial {\bf V}}\cdot \left({\bf V}f_i\right)
=\sum_{j=1}^2J_{ij}^{\text{E}}\left[ {\bf V}|f_{i}(t),f_{j}(t)\right] \;.
\label{3.4} 
\end{equation} 
In the Lagrangian frame, the Enskog collision operator $J_{ij}^{\text{E}}\left[ {\bf V}|f_{i}(t),f_{j}(t)\right] $ becomes
\begin{eqnarray} 
\label{3.5}
J_{ij}^{\text{E}}\left[ {\bf V}_{1}|f_{i},f_{j}\right] &=&\sigma _{ij}^{2}\chi_{ij}\int d{\bf V} 
_{2}\int d\widehat{\boldsymbol {\sigma }}\,\Theta (\widehat{{\boldsymbol {\sigma }}} 
\cdot {\bf g})(\widehat{\boldsymbol {\sigma }}\cdot {\bf g})  \nonumber \\ &&\times \left[ \alpha _{ij}^{-2} f_i({\bf V}_1',t)f_j({\bf V}_2'+a\sigma_{ij}\widehat{\sigma}_y{\widehat{{\bf x}}} ,t)
-f_i( {\bf V}_1,t)f_j({\bf V}_2-a\sigma_{ij}\widehat{\sigma}_y{\widehat{{\bf x}}},t)\right].
\end{eqnarray} 
Here, we have taken into account that $\chi_{ij}$ is uniform in the USF problem. Finally, the expressions for the collisional transfer contribution to the pressure tensor ${\sf P}^{\text{c}}$ and the cooling rate $\zeta$ in the Lagrangian frame are 
\begin{eqnarray}
\label{3.6} 
{\sf P}^{\text{c}}&=&\frac{1}{2}\sum_{i=1}^2\sum_{j=1}^2\frac{m_im_j}{m_i+m_j}\chi_{ij}\sigma_{ij}^3
(1+\alpha_{ij})\int d{\bf V}_1\int d{\bf V}_2\int d\widehat{\boldsymbol {\sigma }}\,\Theta (\widehat{{\boldsymbol {\sigma}}} 
\cdot {\bf g})(\widehat{\boldsymbol {\sigma }}\cdot {\bf g})^2  \nonumber\\
& & \times 
\widehat{\boldsymbol {\sigma }}\widehat{\boldsymbol {\sigma }}
f_i\left({\bf V}_1+a\sigma_{ij}\widehat{\sigma}_y{\widehat{{\bf x}}},t\right)f_j({\bf V}_2,t),
\end{eqnarray}
\begin{eqnarray}
\label{3.7} 
\zeta&=&\frac{1}{6nT}\sum_{i=1}^2\sum_{j=1}^2\frac{m_im_j}{m_i+m_j}\chi_{ij}\sigma_{ij}^2
(1-\alpha_{ij}^2)\int d{\bf V}_1\int d{\bf V}_2\int d\widehat{\boldsymbol {\sigma }}\,\Theta (\widehat{{\boldsymbol {\sigma}}} 
\cdot {\bf g})(\widehat{\boldsymbol {\sigma }}\cdot {\bf g})^3  \nonumber\\
& & \times 
f_i\left({\bf V}_1+a\sigma_{ij}\widehat{\sigma}_y{\widehat{{\bf x}}},t\right)f_j({\bf V}_2,t).
\end{eqnarray}

The normal solution for the USF state adopts the form
\begin{equation}
\label{3.8}
f_i({\bf r}, {\bf v},t)=f_i({\bf V},T(t)), 
\end{equation}
i.e. all the space dependence is accounted for by the flow velocity while all the time dependence appears through the temperature. The Chapman-Enskog method provides this normal solution as an expansion for small spatial gradients, i.e., as a power series in the shear rate $a$:
\begin{equation}
\label{3.9}
f_{i}=f_{i}^{(0)}+f_{i}^{(1)}+\cdots,
\end{equation}
where $f_i^{(k)}$ is of order $k$ in $a$. The time derivatives of the fields, the Enskog collision operator, and the pressure tensor are also expanded as 
\begin{equation}
\label{3.10}
\partial_t=\partial_t^{(0)}+ \partial_t^{(1)}+\cdots,\quad J_{ij}^{\text{E}}=J_{ij}^{(0)}+J_{ij}^{(1)}+\cdots,
\end{equation}
\begin{equation}
\label{3.9bis}
{\sf P}={\sf P}^{(0)}+{\sf P}^{(1)}+\cdots.
\end{equation}
The coefficients in the time derivative expansion are identified by a representation of the momentum flux, the cooling rate, and the external parameter force $\xi$ in the energy balance equation (\ref{3.3}) as a similar series through their definitions as functionals of $f_i$. Consequently, the action of the operator $\partial_t^{(k)}$ is
\begin{equation}
\label{3.10bis}
\partial_t^{(0)}T=-\left(\zeta^{(0)}-\xi^{(0)}\right)T,\quad \partial_t^{(1)}T=0,
\end{equation}
\begin{equation}
\label{3.10bbis}
\partial_t^{(k)}T=-\frac{2}{3n}aP_{xy}^{(k-1)}-(\zeta^{(k)}-\xi^{(k)})T,\quad k\geq 2.
\end{equation}
Upon writing these equations we have taken into account that $P_{xy}^{(0)}=\zeta^{(1)}=\xi^{(1)}=0$. The last equality follows from the fact that the cooling rate is a scalar, and contributions to $\zeta$ in the  first order in the gradients can arise only from $\nabla\cdot {\bf u}$, which is zero in the USF. 

The leading term is the solution to the nonlinear equation 
\begin{equation}
\label{3.11}
\partial_t^{(0)}f_i^{(0)}+\frac{1}{2}\xi^{(0)}\frac{\partial}{\partial {\bf V}}\cdot \left({\bf V}f_i^{(0)}\right)
=\sum_{j=1}^2J_{ij}^{(0)}[f_i^{(0)},f_j^{(0)}],
\end{equation}
where 
\begin{eqnarray}
\label{3.12}
J_{ij}^{(0)}[f_i^{(0)},f_j^{(0)}]&=&\chi_{ij}\sigma_{ij}^2\int d{\bf V}_2 \int 
d\widehat{\boldsymbol {\sigma }}\,\Theta (\widehat{{\boldsymbol {\sigma}}} 
\cdot {\bf g})(\widehat{\boldsymbol {\sigma }}\cdot {\bf g}) \nonumber\\
& & \times 
\left[\alpha_{ij}^{-2}f_i^{(0)}({\bf V}_1')f_j^{(0)}({\bf V}_2')-f_i^{(0)}({\bf V}_1)f_j^{(0)}({\bf V}_2)\right].
\end{eqnarray}
Dimensional analysis requires that $f_i^{(0)}({\bf V})$ must be of the form 
\begin{equation}
\label{3.14}
f_i^{(0)}(V)=n_iv_0^{-3}\Phi_i(V/v_0)
\end{equation}
where 
\begin{equation}
\label{3.15}
v_0=\sqrt{2T\sum_{i=1}^2 m_i^{-1}}
\end{equation}
is a thermal velocity defined in terms of the temperature $T$ of the 
mixture. According to (\ref{3.14}), the time derivative in (\ref{3.11}) can be represented more usefully as 
\begin{equation}
\label{3.16}
\partial_t^{(0)}f_i^{(0)}=-(\zeta^{(0)}-\xi^{(0)})T\partial_Tf_i^{(0)}=\frac{1}{2}(\zeta^{(0)}-\xi^{(0)})\frac{\partial}{\partial {\bf V}}\cdot \left({\bf V}f_i^{(0)}\right).
\end{equation}
The Enskog equation at this order can be written finally as 
\begin{equation}
\label{3.16bis}
 \frac{1}{2}\zeta^{(0)}\frac{\partial}{\partial {\bf V}}\cdot \left({\bf V}f_i^{(0)}\right)
=\sum_{j=1}^2J_{ij}^{(0)}[f_i^{(0)},f_j^{(0)}].
\end{equation}
Therefore, Eq.\ (\ref{3.11}) happens to be  formally {\em identical} to the one obtained in the unforced case (i.e., with $\xi=0$) \cite{GD99b}, and consequently there is an exact correspondence between the homogeneous cooling state and this type of driven steady state. This is one of the advantages of the Gaussian thermostat. Since the distribution functions $f_i^{(0)}$ are isotropic, the zeroth order pressure tensor is found from Eqs.\ (\ref{2.14}) and (\ref{3.6}) as $P_{\alpha\beta}^{(0)}=p\delta_{\alpha\beta}$, where the pressure $p$ is 
\begin{eqnarray}
\label{3.18}
p&=&\sum_{i=1}^2n_iT_i+\frac{1}{6}\sum_{i=1}^2\sum_{j=1}^2\frac{m_im_j}{m_i+m_j}\sigma_{ij}^3
\chi_{ij}(1+\alpha_{ij})\int d{\bf V}_1\int d{\bf V}_2 f_i^{(0)}(V_1)f_j^{(0)}(V_2)\nonumber\\
& & \times
\int d\widehat{\boldsymbol {\sigma }}\,\Theta (\widehat{{\boldsymbol {\sigma}}} 
\cdot {\bf g})(\widehat{\boldsymbol {\sigma }}\cdot {\bf g})^2  \nonumber\\
&=&\sum_{i=1}^2n_iT_i+\frac{2\pi}{3}\sum_{i=1}^2\sum_{j=1}^2\sigma_{ij}^3\chi_{ij}n_in_j
\mu_{ji}(1+\alpha_{ij})T_i.
\end{eqnarray}
Here, we have introduced the kinetic  temperatures $T_i$ for each species defined as 
\begin{equation}
\label{2.12}
\frac{3}{2}n_iT_i=\int d{\bf v}\frac{m_{i}}{2}V^{2}f_{i}^{(0)}.
\end{equation}
As said in the Introduction, in general the partial temperatures $T_i$ differ from the (global) temperature $T$ and so the total energy is not equally distributed between both species (breakdown of energy equipartition). 

The analysis to first order in $a$ is worked out in Appendix \ref{appB}. The distribution $f_1^{(1)}$ obeys the integral equation
\begin{equation}
\label{3.19}
\left[(\xi^{(0)}-\zeta^{(0)})T\partial_T+\frac{1}{2}\xi^{(0)}\frac{\partial}{\partial {\bf V}}\cdot {\bf V}+
{\cal L}_1\right]f_1^{(1)}+{\cal M}_1f_2^{(1)}=aV_y\frac{\partial}{\partial V_x} 
f_{1}^{(0)}+a\sum_{j=1}^{2} \Lambda_{1j} [f_1^{(0)},f_j^{(0)}].
\end{equation}
A similar equation can be obtained for $f_2^{(1)}$, by just making the changes $1\leftrightarrow 2$. The specific form of the linear operators ${\cal L}_i, {\cal M}_i$, and $\Lambda_{ij}$ are also given in Appendix \ref{appB}. The contributions  $f_i^{(0)}$ and $f_i^{(1)}$ determine the pressure tensor ${\sf P}^{(1)}$ to first order in the shear rate. The result is 
\begin{equation}
\label{3.20}
P_{\alpha\beta}^{(1)}=-\eta a\left( \delta_{\alpha x}\delta_{\beta y}+\delta_{\alpha y}\delta_{\beta x}\right),
\end{equation}
where $\eta$ is the shear viscosity coefficient. This coefficient has kinetic and collisional transfer contributions
\begin{equation}
\label{3.21}
\eta=\eta^{\text{k}}+\eta^{\text{c}}.
\end{equation}
The kinetic contribution $\eta^{\text{k}}$ is given by 
\begin{equation}
\label{3.22}
\eta^{\text{k}}=\sum_{i=1}^2 \eta_i^{\text{k}}, \quad  \eta_i^{\text{k}}=-\frac{m_i}{a}\int d{\bf V} V_xV_y f_i^{(1)}({\bf V}),
\end{equation}
while the collisional contribution $\eta^{\text{c}}$ is
\begin{equation}
\label{3.23}
\eta^{\text{c}}=\frac{4\pi}{15}\sum_{i=1}^2\sum_{j=1}^2\sigma_{ij}^3\chi_{ij}(1+\alpha_{ij})n_j\mu_{ji}\left[\eta_i^{\text{k}}+
\frac{m_i\sigma_{ij}}{4n_j}\int d{\bf V}_1\int d{\bf V}_2f_i^{(0)}({\bf V}_1)f_j^{(0)}({\bf V}_2)g\right].
\end{equation}

\section{Sonine polynomial approximation}
\label{sec3bis}

For practical purposes the integral equations (\ref{3.16bis}) and (\ref{3.19}) for $f_i^{(0)}$ and $f_i^{(1)}$ are solved by using low order truncation of expansions in a series of Sonine polynomials. The polynomials are defined with respect to a Gaussian weight factor whose parameters are chosen such that the leading term in the expansion yields the exact moments of the entire distribution with respect to $1$, $m_i {\bf v}$ and $\case{1}{2}m_iv^2$.  In the leading order, the distribution $\Phi_i$ appearing in Eq.\ (\ref{3.14}) is given by 
\begin{equation}
\label{3.17}
\Phi_i(V^*)\to \left(\frac{\theta_i}{\pi}\right)^{3/2}e^{-\theta_iV^{*2}}
\left[1+\frac{c_i}{4}\left(\theta_iV^{*4}-5\theta_iV^{*2}+\frac{15}{4}\right)
\right],
\end{equation}
where $V^*=V/v_0$, 
\begin{equation}
\label{3.17bis}
\theta_i=\frac{m_i}{\gamma_i}\sum_{j=1}^2m_j^{-1},
\end{equation}
and $\gamma_i=T_i/T$. For elastic collisions, $\gamma_i=1$, i.e., the partial temperatures $T_i$ coincide with the global temperature $T$. In the inelastic case, $\gamma_i\neq 1$ and presents a complex dependence on the parameters of the problem.  The coefficients $c_i$ (which measure the deviation of $\Phi_i$ from the reference Maxwellian) are determined consistently from the Enskog equation. The approximation (\ref{3.17}) provides detailed predictions for the cooling rate $\zeta^{(0)}$, the temperature ratio $T_1/T_2$ and the cumulants $c_i$ as functions of the mass ratio, size ratio, composition, density, and restitution coefficients\cite{GD99b}. Recently, the accuracy of this approximate solution has been confirmed by Monte Carlo \cite{MG02} and molecular dynamics simulations \cite{DHGD02} over a wide range of values in the parameter space. 

In the case of the distributions $f_i^{(1)}$, the leading Sonine approximation is
\begin{equation}
\label{3.24}
f_i^{(1)}\to -af_{i,M}\frac{m_i\eta_i^{\text{k}}}{n_iT_i^2}V_xV_y, \quad 
f_{i,M}({\bf V})=n_i(m_i/2T_i)^{3/2}\exp(-m_iV^2/2T_i).
\end{equation}
By using (\ref{3.24}), the partial kinetic contributions $\eta_i^{\text{k}}$ to the shear viscosity can be obtained from Eq.\ (\ref{3.19}) by multiplying it with $m_i V_xV_y$ and integrating over the velocity. From dimensional analysis $\eta_i^{\text{k}}\propto T^{1/2}$ and so one gets the coupled set of equations
\begin{equation}
\label{3.25}
\left(
\begin{array}{cc}
\tau_{11}-\case{1}{2}(\xi^{(0)}+\zeta^{(0)})&\tau_{12}\\
\tau_{21} & \tau_{22}-\case{1}{2}(\xi^{(0)}+\zeta^{(0)})
\end{array}
\right)
\left(
\begin{array}{c}
\eta_1^{\text{k}}/n_1T_1^2\\
\eta_2^{\text{k}}/n_2T_2^2
\end{array}
\right)=
\left(
\begin{array}{c}
T_1^{-1}-\widetilde{\Lambda}_{11}-\widetilde{\Lambda}_{12}\\
T_2^{-1}-\widetilde{\Lambda}_{21}-\widetilde{\Lambda}_{22}
\end{array}
\right),
\end{equation}
where 
\begin{equation}
\label{3.26} 
\tau_{ii}=\frac{1}{n_iT_i^2}\int d{\bf V} m_i V_x V_y {\cal L}_i (f_i^{(1)}), 
\end{equation} 
\begin{equation}
\label{3.27}
\tau_{ij}=\frac{1}{n_iT_i^2}\int d{\bf V} m_i V_x V_y {\cal M}_i (f_j^{(1)}), \quad (i\neq j)
\end{equation} 
\begin{equation}
\label{3.28}
\widetilde{\Lambda}_{ij}=\frac{1}{n_iT_i^2}\int d{\bf V} m_iV_xV_y\Lambda_{ij}[f_i^{(0)},f_j^{(0)}].
\end{equation}
The integrals (\ref{3.28}) are evaluated in Appendix \ref{appC}, while the collision integrals (\ref{3.26}) and (\ref{3.27}) were already evaluated in the Boltzmann limit (except for the factor $\chi_{ij}$). The explicit form of these integrals are also quoted in Appendix \ref{appC}.
The solution of (\ref{3.25}) with the matrix elements known is elementary and so the kinetic contribution $\eta^{\text{k}}$ to the shear viscosity can be easily calculated from Eq.\ (\ref{3.22}). Finally, use of Eq.\ (\ref{3.17}) in Eq.\ (\ref{3.23}) determines the collisional transfer contribution to the shear viscosity. The result is (see Appendix \ref{appC})
\begin{eqnarray}
\label{3.29}
\eta^{\text{c}}=\frac{4\pi}{15}\sum_{i=1}^2\sum_{j=1}^2 & & \sigma_{ij}^3\chi_{ij}(1+\alpha_{ij})n_j 
\mu_{ji}\left\{\eta_i^{\text{k}}+
m_in_i\sigma_{ij}\left(\frac{m_iT_j+m_jT_i}{2\pi m_im_j}\right)^{1/2}\right.\nonumber\\
& & \left.\times \left[1-\frac{c_i}{8}
\left(\frac{m_jT_i}{m_iT_j+m_jT_i}\right)^2\right]\right\}.
\end{eqnarray}

Equations (\ref{3.22}), (\ref{3.25}), and (\ref{3.29}) provide the explicit expression for the shear viscosity $\eta$ of a dense granular binary mixture under driven USF in the first Sonine approximation. This coefficient is given in terms of the restitution coefficients $\alpha_{11}$, $\alpha_{22}$, and $\alpha_{12}$, the temperature $T$, and the parameters of the mixture, namely, the masses $m_i$, the sizes $\sigma_i$, the mole fractions $x_i$ and the solid volume fraction $\phi=\phi_1+\phi_2$. Here, $\phi_i=(\pi/6)n_i\sigma_i^3$ is the species volume fraction of the component $i$. To get the explicit dependence of $\eta$ on $\phi$, the form of the pair correlation function $\chi_{ij}$ at contact must be chosen.  A good approximation for $\chi_{ij}$ for a mixture of hard spheres is given by the generalized Carnahan-Starling form \cite{CS}
\begin{equation}
\label{3.29bis}
\chi_{ij}=\frac{1}{1-\phi}+\frac{3}{2}\frac{\beta}{(1-\phi)^2}\frac{\sigma_i\sigma_j}{\sigma_{ij}} 
+\frac{1}{2}\frac{\beta^2}{(1-\phi)^3}\left(\frac{\sigma_i\sigma_j}{\sigma_{ij}}\right)^2,
\end{equation}
where $\beta=\pi(n_1\sigma_1^2+n_2\sigma_2^2)/6$.

Before studying the general dependence of $\eta$ on the parameter space, let us consider some special limit cases. In the elastic limit, $\alpha_{11}=\alpha_{22}=\alpha_{12}=1$, $\zeta=0$, $\gamma_i=1$, $\theta_1=1/\mu_{21}$, $\theta_2=1/\mu_{12}$, and $c_1=c_2=0$. In this case, the shear viscosity coefficient of an unforced ($\xi=0$) mixture can be written as 
\begin{equation}
\label{3.31}
\eta=\sum_{i=1}^2\left(1+\frac{8\pi}{15}\sum_{j=1}^2\sigma_{ij}^3\chi_{ij}n_j\mu_{ji}\right)\eta_i^{\text{k}}
+\frac{4}{15}\sum_{i=1}^2\sum_{j=1}^2 \left(\frac{2\pi m_im_jT}{m_i+m_j}\right)^{1/2}n_in_j\sigma_{ij}^4
\chi_{ij},
\end{equation}
where now the kinetic contributions $\eta_i^{\text{k}}$ verify the set of equations (\ref{3.25}) with  $\zeta^{(0)}=\xi^{(0)}=0$ and 
\begin{equation}
\label{3.30}
\tau_{ij}=\frac{16}{15}\sum_{\ell=1}^2 \frac{n_\ell \sigma_{i\ell}^2\chi_{i\ell}}{(m_i+m_\ell)^{3/2}}
\left(\frac{2\pi Tm_\ell}{m_i}\right)^{1/2}\left( 5 m_i\delta_{ij}+3m_\ell \delta_{ij}-2m_i\delta_{j\ell}\right), 
\end{equation}
\begin{equation}
\label{3.32}
\widetilde{\Lambda}_{ij}=-\frac{8\pi}{15}\frac{n_j\sigma_{ij}^3}{T}\mu_{ji}\chi_{ij}.
\end{equation}
Equations (\ref{3.31})--(\ref{3.32}) agree with the first Sonine approximation to the coefficient of shear viscosity of a molecular gas-mixture of hard spheres\cite{TG71}.  In the case of mechanically equivalent (inelastic) particles, $\gamma_i=1$, $\zeta_1=\zeta_2\equiv \zeta$ and $c_1=c_2\equiv c$, where 
\begin{equation}
\label{3.33}
\zeta=\frac{4}{3}n\sigma^2 \sqrt{\frac{\pi T}{m}}(1-\alpha^2)\left(1+\frac{3c}{32}\right),
\end{equation}
\begin{equation}
\label{3.34}
c=\frac{32(1-\alpha)(1-2\alpha^2)}{81-17\alpha+30\alpha^2(1-\alpha)}.
\end{equation}
In this case, Eqs.\ (\ref{3.22}), (\ref{3.25}), and (\ref{3.29}) yield
\begin{equation}
\label{3.35}
\eta=\eta^{\text{k}}\left[1+\frac{4 \phi \chi (1+\alpha)}{5}\right]+\frac{4}{5}\sqrt{\frac{nT}{\pi}}\phi n \chi \sigma (1+\alpha)(1-\frac{c}{32}),
\end{equation}
and the kinetic part $\eta^{\text{k}}$ is
\begin{equation}
\label{3.36}
\eta^{\text{k}}=\frac{nT}{\nu_{\eta}-\case{1}{2}(\xi^{(0)}+\zeta^{(0)})}\left[1-\frac{2}{5}(1+\alpha)(1-3\alpha)\phi \chi\right],
\end{equation}
where
\begin{equation}
\label{3.37}
\nu_\eta=\frac{16}{5}n\sigma^2\sqrt{\frac{\pi T}{m}}\chi\left[1-\frac{1}{4}(1-\alpha)^2\right]\left(1-\frac{c}{64}\right).
\end{equation}
The expression (\ref{3.35}) coincides with the one recently obtained for a granular monocomponent gas \cite{GM02,GD99a}. Finally, when $\phi\to 0$ it is easy to check that the results derived here reduce to those previously found in Ref.\ \cite{MG03} for a dilute gas. This shows the self-consistency of the present description.

\begin{figure}
\includegraphics[width=0.5 \columnwidth]{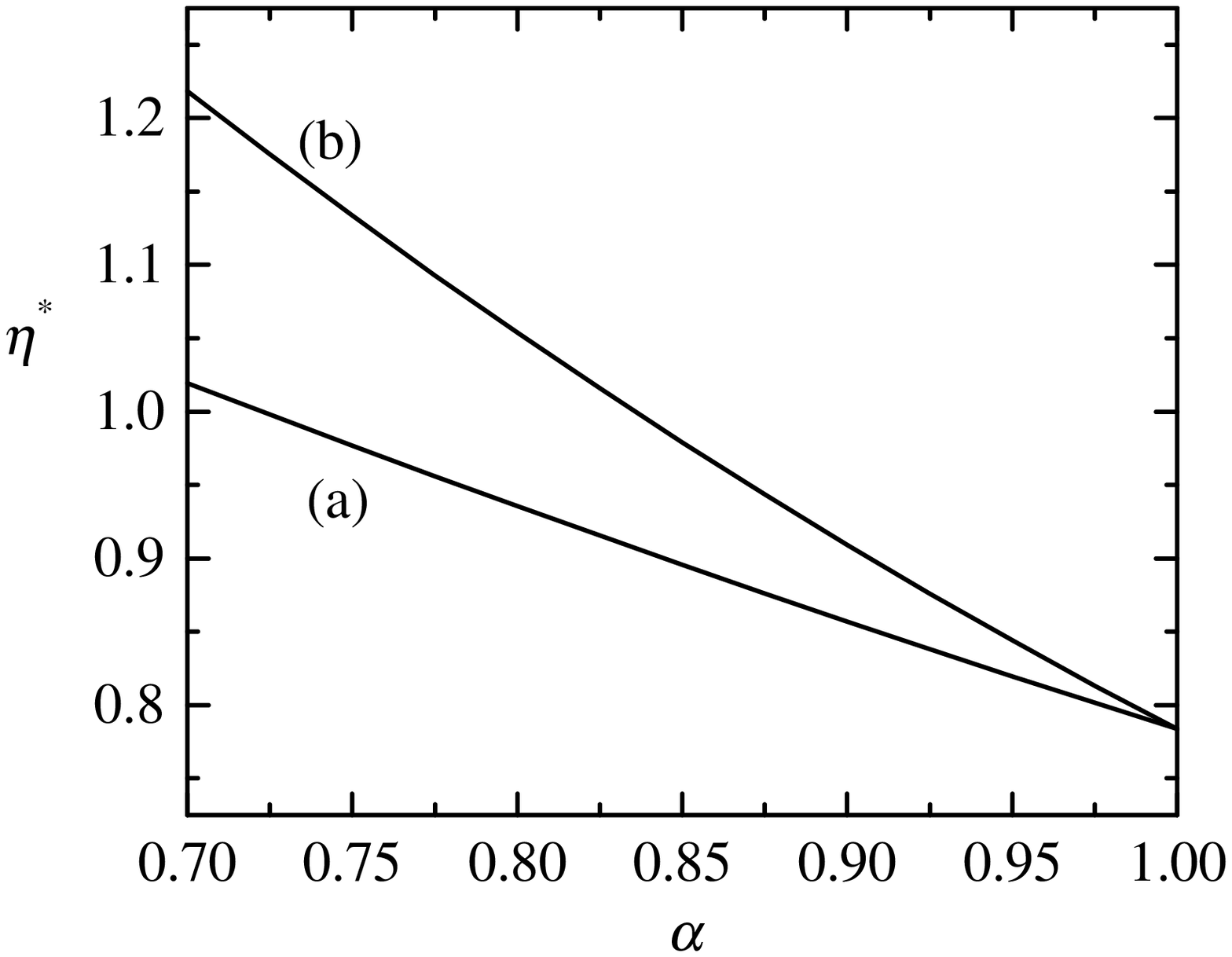}
\caption{Plot of the reduced shear viscosity $\eta^*$ as a function of the restitution coefficient $\alpha$ for a binary mixture with parameters  $\phi=0$, $x_1=1/2$, $\sigma_1/\sigma_2=1$, and $m_1/m_2=4$ in (a) the unforced case ($\xi^{(0)}=0$) and (b) the forced case ($\xi^{(0)}=\zeta^{(0)})$.
\label{fig1}}
\end{figure}

Before comparing the kinetic theory predictions with numerical simulation data, it is instructive to compare the results obtained in the unforced ($\xi^{(0)}=0$) and driven ($\xi^{(0)}=\zeta^{(0)}$) cases. In Fig.\  \ref{fig1} we plot the reduced shear viscosity $\eta^*$ as a function of the (common) restitution coefficient $\alpha_{ij}=\alpha$ for  $\sigma_1/\sigma_2=1$, $m_1/m_2=4$, $x_1=1/2$, and $\phi=0$ in the above two cases. Here, the reduced shear viscosity $\eta^*$ is defined as 
\begin{equation}
\label{3.38}
\eta^*=\frac{\nu}{nT}\eta,
\end{equation}
where 
\begin{equation}
\label{3.39}
\nu=\sqrt{\pi}n\sigma_{12}^2v_0
\end{equation}
is an effective collision frequency. We see that the Navier-Stokes shear viscosity of the (unforced) gas differs from the shear viscosity of the gas when the latter is {\em excited} by the (Gaussian) external force, the discrepancy increasing as the restitution coefficient decreases. This shows again that the driving force does not play a neutral role in the problem and the transport property is affected by this type of external forcing mechanism\cite{GM02}.  However, for practical purposes, the introduction of these driving forces has the advantage of that they can be incorporated into the kinetic theory very easily and they allow, for instance, to test the validity of some of the underlying assumptions made in the theory through a direct comparison with computer simulations.

\section{Monte Carlo simulation for uniform shear flow}
\label{sec4}

The expression obtained in the previous section for the shear viscosity requires the truncation of an expansion of the integral equations in Sonine polynomials. To assess the degree of accuracy of this approximation, one has to resort to numerical solutions of the Enskog equation, such as those obtained from Monte Carlo simulations. In this section, we briefly describe the method employed in the simulation in the case of the USF state. 

For a granular fluid under USF and in the absence of a thermostating force ($\xi=0$), the energy balance (\ref{3.3}) leads to a steady state when the viscous heating effect is exactly balanced by the collisional cooling\cite{MG02bis}. However, when the granular mixture is excited by the Gaussian force
\begin{equation}
\label{4.3}
{\bf F}_i=\frac{1}{2}m_i\xi{\bf V}, 
\end{equation}
that exactly compensates for the collisional energy loss ($\xi=\zeta$), the viscous heating dominates and the temperature obeys the equation 
\begin{equation}
\label{4.3bis}
\frac{\partial T}{\partial t}=-\frac{2}{3n}aP_{xy}.
\end{equation}
Since the granular temperature $T$ increases in time, so does the collision frequency $\nu(t)\propto \sqrt{T(t)}$, and hence the reduced shear rate $a^*(t)=a/\nu(t)$ (which is the relevant uniformity parameter) monotonically decreases in time. Under these conditions, the system asymptotically reaches a regime described by linear hydrodynamics and the (reduced) Navier-Stokes shear viscosity $\eta^*$ can be measured as\cite{NO79} 
\begin{equation}
\label{4.2}
\eta^*=-\lim_{t\to \infty}\frac{P_{xy}^*}{a^*},
\end{equation}
where $P_{xy}^*=P_{xy}/nT$. Recently, this idea has been used to identify the shear viscosity of a (heated) granular binary mixture in the low-density regime \cite{MG03}. The comparison with kinetic theory showed an excellent agreement over a wide range of values of the restitution coefficient and the rest of parameters characterizing the system.

We have numerically solved Eq.\ (\ref{3.4}) by means of an extension of the well-known Direct Simulation Monte Carlo (DSMC) method \cite{B94} to dense gases. The method is usually referred to as the Enskog Simulation Monte Carlo (ESMC) method. This method was devised to mimic the dynamics involved in the Enskog collision term and it has been previously used to analyze rheological properties of a elastic dense gas\cite{MS96} and the shock-wave structure\cite{MLGS98}. In the present work, the ESMC algorithm has been modified to study the dynamics of a granular binary mixture of a finite density. Since the USF  is spatially homogeneous in the local Lagrangian frame, the simulation method becomes especially easy to implement and efficient from a computational point of view. This is an important advantage with respect to molecular dynamics simulations. Nevertheless, the restriction to this homogeneous state prevents us from analyzing the possible instability of USF or the formation of clusters or microstructures. 

The ESMC method as applied to a granular binary mixture under USF is as follows. The velocity distribution function of the species $i$ is represented by the peculiar velocities $\{{\bf V}_k\}$ of $N_i$ ``simulated" particles:
\begin{equation}
\label{4.1}
f_i({\bf V},t)\to n_i \frac{1}{N_i}\sum_{k=1}^{N_i} \delta({\bf V}-{\bf 
V}_k(t))\; .
\end{equation}
Note that the number of particles $N_i$ is arbitrary, but must be taken according to the relation $N_1/N_2=n_1/n_2$. At the initial state, one assigns velocities to the particles drawn from the Maxwell-Boltzmann probability distribution: 
\begin{equation}
\label{4.2b}
f_i({\bf V},0)=n_i\ \pi^{-3}\ V_{0i}^{-3}(0)\ \exp\left(-V^2/V_{0i}^2(0)\right)\;,
\end{equation}
where $V_{0i}^2(0)=2T(0)/m_i$ and $T(0)$ is the initial temperature. To enforce a vanishing initial total momentum, the velocity of every particle is subsequently subtracted by the amount $N_i^{-1} \sum_k {\bf V}_k(0)$. In the ESMC method, the free motion and the collisions are uncoupled over a time step $\Delta t$ which is small compared with both the mean free time and the inverse shear rate. As $a^*$ decreases monotonically in time, the value of $\Delta t$ must be updated in the course of the simulation. In the local Lagrangian frame, particles of each species ($i=1,2$) are subjected to the action of a non-conservative inertial force ${\bf F}_i=-m_i\ {\sf a}\cdot{\bf V}$. Thus, the free motion stage consists of making ${\bf V}_k\to {\bf V}_k-{\sf a}\cdot{\bf V}_k\Delta t$. In the collision stage, binary interactions between particles of species $i$ and $j$ must be considered. To simulate the collisions between particles of species $i$ with $j$ a sample of $\frac{1}{2} N_i \omega_{\text{max}}^{(ij)}\Delta t$ pairs is chosen at random with equiprobability. Here, $\omega_{\text{max}}^{(ij)}$ is an upper bound estimate of the probability that a particle of the species $i$ collides with a particle of the species $j$. Let us consider a pair $(k,\ell)$ belonging to this sample. Hereafter, $k$ denotes a particle of species $i$ and $\ell$ a particle of species $j$. For each pair $(k,\ell)$ with velocities $({\bf V}_k,{\bf V}_{\ell})$, the following steps are taken: (1) a given direction $\widehat{\boldsymbol \sigma}_{k\ell}$ is chosen at random with equiprobability; (2) the collision between particles $k$ and $\ell$ is accepted with a probability equal to $\Theta({\bf g}_{k\ell}\cdot \widehat{\boldsymbol \sigma}_{k\ell})\omega_{k\ell}^{(ij)}/ \omega_{\text{max}}^{(ij)}$, where $\omega_{k\ell}^{(ij)}=4\pi \sigma_{ij}^2 n_j|{\bf g}_{k\ell}\cdot \widehat{\boldsymbol \sigma}_{k\ell}|$ and ${\bf g}_{k\ell}={\bf V}_k-{\bf V}_{\ell}-\sigma_{ij} {\sf a}\cdot \widehat{\boldsymbol \sigma}_{k\ell}$; (3) if the collision is accepted, postcollisional velocities are assigned to both particles according to the scattering rules:
\begin{equation}
\label{4.3b}
{\bf V}_{k}\to {\bf V}_{k}-\mu_{ji}(1+\alpha_{ij})({\bf g}_{k\ell}\cdot \widehat{\boldsymbol 
\sigma}_{k\ell})\widehat{\boldsymbol \sigma}_{k\ell}\; ,
\end{equation}
\begin{equation}
\label{4.4b}
{\bf V}_{\ell}\to {\bf V}_{\ell}+\mu_{ij}(1+\alpha_{ij})({\bf g}_{k\ell}\cdot \widehat{\boldsymbol 
\sigma}_{k\ell})\widehat{\boldsymbol \sigma}_{k\ell}\;.
\end{equation}
If in a collision $\omega_{k\ell}^{(ij)}>\omega_{\text{max}}^{(ij)}$, the estimate of $\omega_{\text{max}}^{(ij)}$ is updated as $\omega_{\text{max}}^{(ij)}=\omega_{k\ell}^{(ij)}$. The procedure described above is performed for $i=1,2$ and $j=1,2$. The granular temperature is calculated before and after the collision stage, and thus the instantaneous value of the cooling rate $\zeta$ is obtained. After the collisions have been calculated, the thermostat force (\ref{4.3}) is considered by making ${\bf V}_k\to {\bf V}_k+1/2\ \zeta {\bf V}_k\Delta t$.

In the course of the simulations, one evaluates the kinetic and collisional transfer contributions to the pressure tensor. They are given as 
\begin{equation}
\label{4.5}
{\sf P}^{\text{k}}=\sum_{i=1}^{2} \frac{m_i n_i}{N_i}\sum_{k=1}^{N_i} {\bf V}_k {\bf V}_k\; ,
\end{equation}
\begin{equation}
{\sf P}^{\text{c}}=\frac{n}{2N\Delta t}{\sum_{k\ell}}^{\dagger} \mu_{ij}m_j \sigma_{ij}(1+\alpha_{ij})
({\bf g}_{k\ell}\cdot \widehat{\boldsymbol \sigma}_{k\ell})\widehat{\boldsymbol \sigma}_{k\ell}
\widehat{\boldsymbol \sigma}_{k\ell}\; ,
\end{equation}
where the dagger means that the summation is restricted to the accepted collisions. The shear viscosity $\eta$ is obtained from (\ref{4.2}). To improve the statistics, the results are averaged over a number ${\cal N}$ of independent realizations or replicas. In our simulations we have typically taken a total number of particles $N=N_1+N_2=10^5$, a number of replicas ${\cal N}=10$, and a time step $\Delta t=3\times 10^{-3} \lambda_{11}/V_{01}(0)$, where $\lambda_{11}=(\sqrt{2} \pi n_1 \sigma_{11}^2)^{-1}$ is the mean free path for collisions 1--1.

\begin{figure}
\includegraphics[width=0.5 \columnwidth]{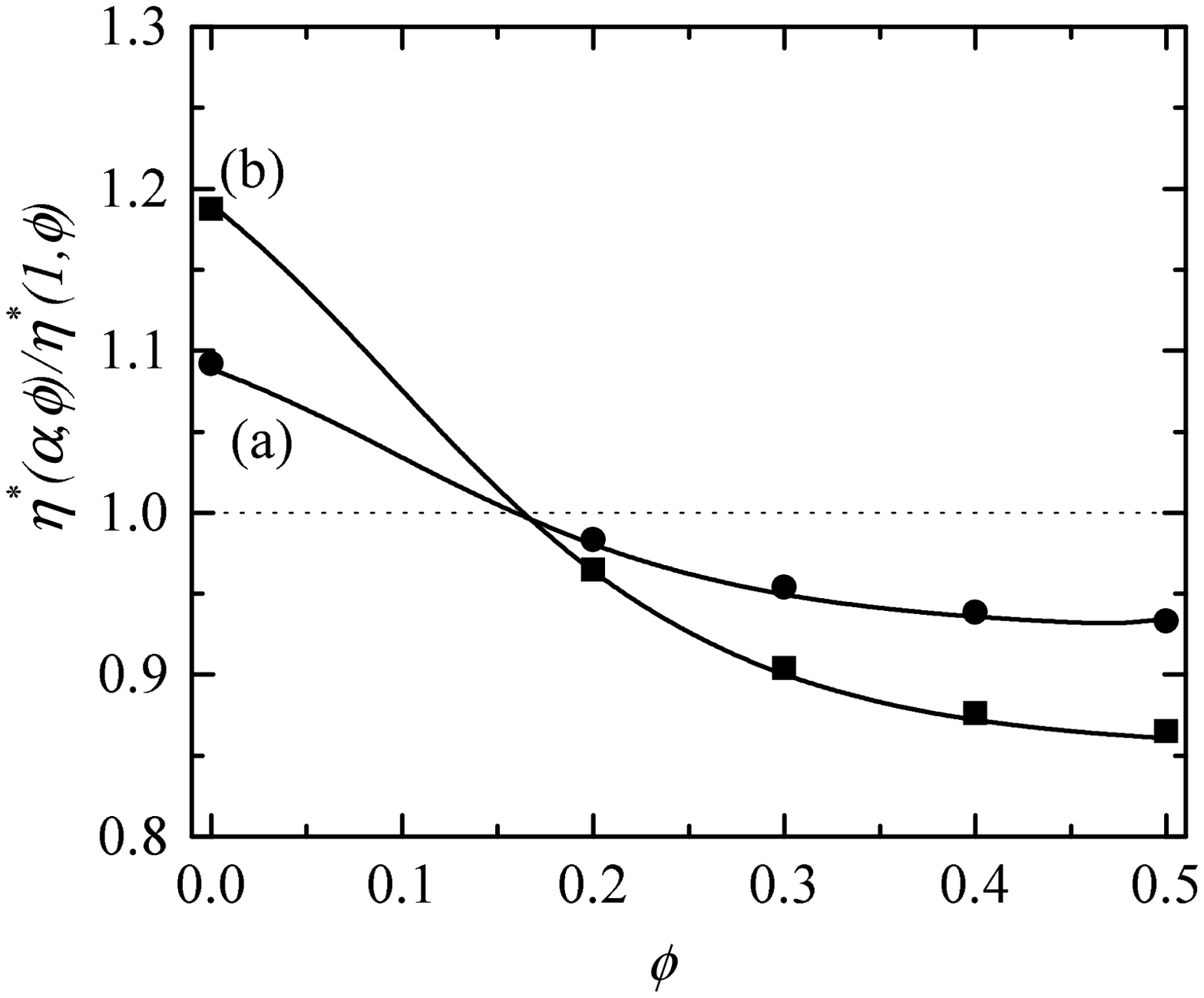}
\caption{Plot of the ratio $\eta^*(\alpha,\phi)/\eta^*(1,\phi)$ for a monocomponent gas as a function of the solid fraction $\phi$ for two different values of the restitution coefficient $\alpha$: (a) $\alpha=0.9$ (circles), and (b) $\alpha=0.8$ (squares). The lines are the theoretical predictions and the symbols refer to the results obtained from Monte Carlo simulations.
\label{fig2}}
\end{figure}

\section{Results}
\label{sec5}

In this section we compare the results obtained from the Chapman-Enskog method for the shear viscosity coefficient of a {\em heated} granular mixture (i.e., with $\xi^{(0)}=\zeta^{(0)}$) with those obtained from the ESMC method.  For the sake of simplicity, we assume that $\alpha_{11}=\alpha_{22}=\alpha_{12}\equiv \alpha$ so that we reduce the parameter set of the problem to five quantities: $\{\alpha, \phi, m_1/m_2, \sigma_1/\sigma_2, x_1\}$.  For concreteness, henceforth we will assume that $m_1\geq m_2$ and $\sigma_1\geq \sigma_2$. To compare and contrast the results of a binary mixture with that of its monocomponent counterpart, we first show some results for a monodisperse system over a range of solid fractions and restitution coefficients. 
\begin{figure}
\includegraphics[width=0.5 \columnwidth]{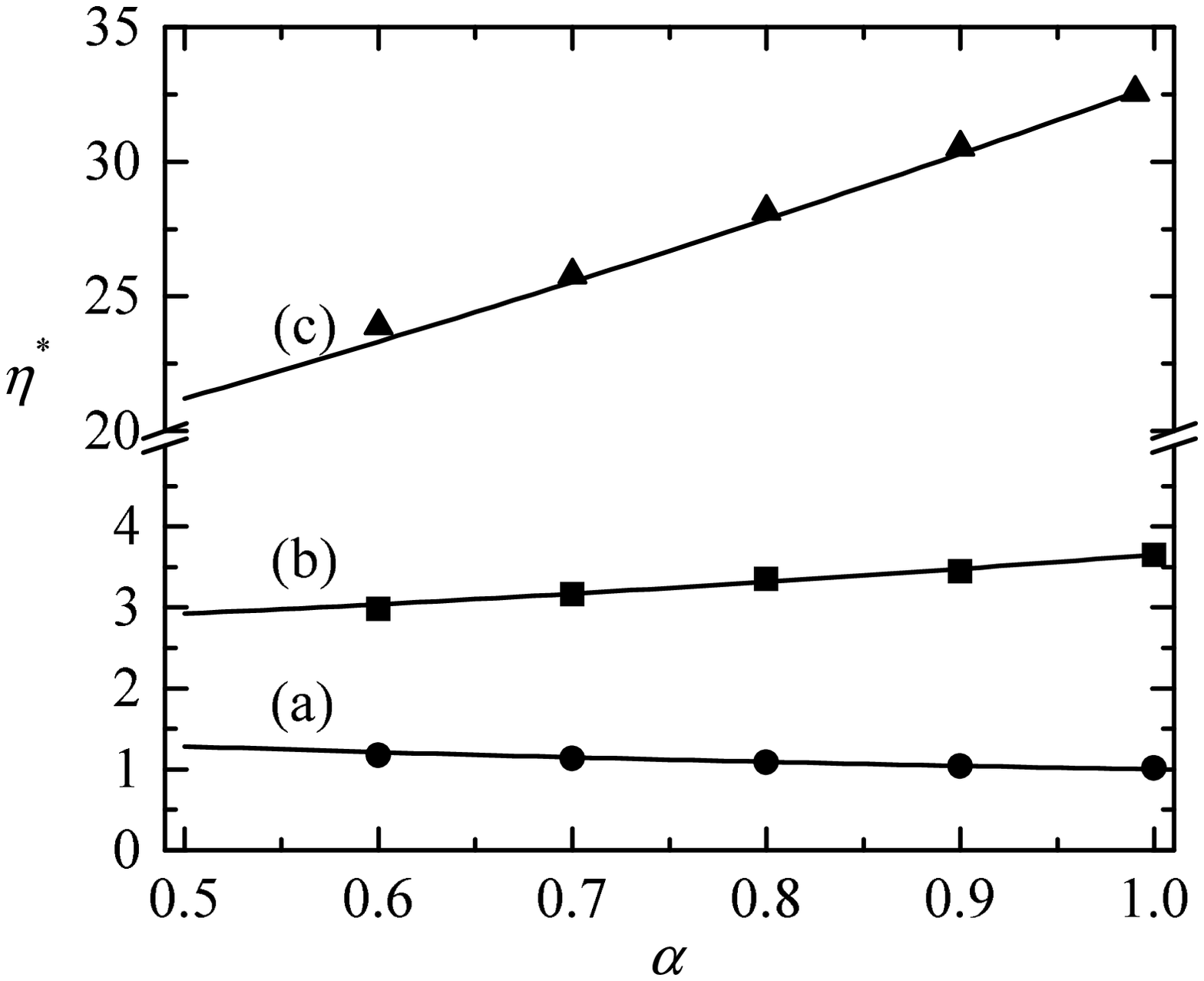}
\caption{Plot of the reduced shear viscosity $\eta^*$ of a monocomponent gas as a function of the restitution coefficient $\alpha$ for three different values of the solid fraction $\phi$: (a) $\phi=0$ (circles), (b) $\phi=0.2$ (squares), and (c) $\phi=0.4$ (triangles). The lines are the theoretical predictions and the symbols refer to the results obtained from Monte Carlo simulations.
\label{fig3}}
\end{figure}

\subsection{Monocomponent dense gas}

Figure \ref{fig2} shows the dependence of the ratio  $\eta^*(\alpha,\phi)/\eta^*(1,\phi)$ on the 
solid fraction $\phi$ for two values of the restitution coefficient. The symbols represent the simulation data while the lines refer to the theoretical results obtained from the Enskog equation, Eqs.\ (\ref{3.35})--(\ref{3.37}). Both theory and simulation show that, for a given value of the density,  the shear viscosity increases with decreasing $\alpha$ (i.e., greater dissipation) {\em if} the solid fraction is smaller than a threshold value $\phi_0(\alpha)$, while the opposite happens if $\phi>\phi_0(\alpha)$. Similar threshold values exist for the kinetic and collisional parts of the shear viscosity. We observe that in the range $0.8\leq\alpha \leq 1$, the kinetic theory calculations show that these threshold values are practically independent of the restitution coefficient. Specifically, $\phi_0(\alpha)\simeq 0.16$, while the corresponding values for the kinetic and collisional parts are, respectively, 0.23 and  0.05. It is apparent that the comparison between Monte Carlo simulation data and theoretical results shows an excellent agreement over the entire range of densities considered. The dependence of $\eta^*(\alpha,\phi)$ on dissipation is plotted in Fig.\ \ref{fig3} for three different values of the solid fraction. We see that in general the influence of dissipation on the shear viscosity $\eta^*(\alpha)$ is quite significant, except for $\phi=0.2$ which is very close to the threshold value $\phi_0$.
As in Fig.\ \ref{fig2}, the theory compares quite well with simulation data, except perhaps at  $\phi=0.4$ for strong dissipation ($\alpha=0.6$).

\begin{figure}
\includegraphics[width=0.5 \columnwidth]{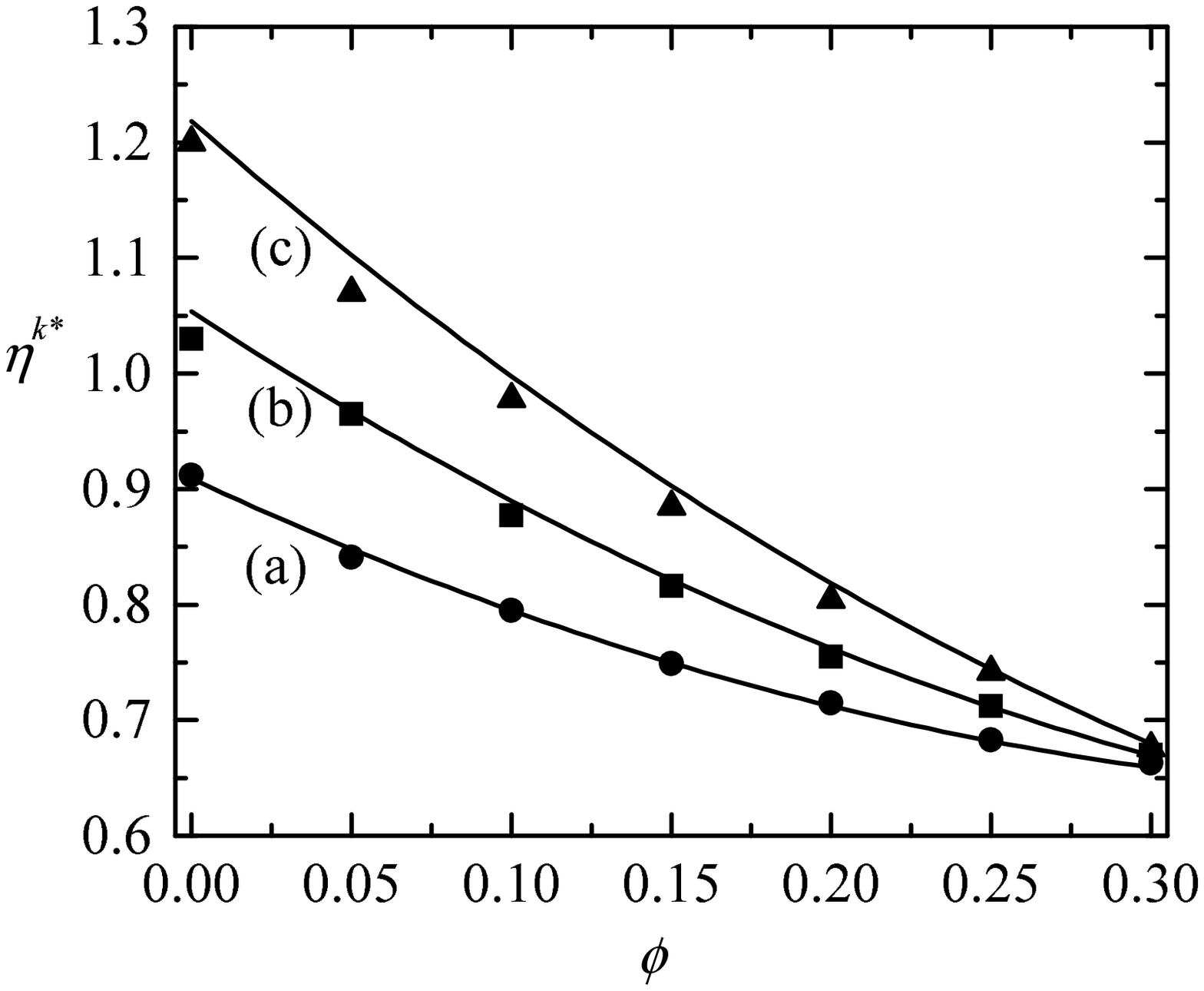}
\caption{Plot of the kinetic part $\eta^{\text{k}*}$ of the reduced shear viscosity as a function of the solid fraction $\phi$ for $m_1/m_2=4$, $\sigma_1/\sigma_2=1$, $x_1=1/2$ and three different values of the restitution coefficient $\alpha$: (a) $\alpha=0.9$ (circles), (b) $\alpha=0.8$ (squares), and (c) $\alpha=0.7$ (triangles). The lines are the theoretical predictions and the symbols refer to the results obtained from Monte Carlo simulations.
\label{fig4}}
\end{figure}

\begin{figure}
\includegraphics[width=0.5 \columnwidth]{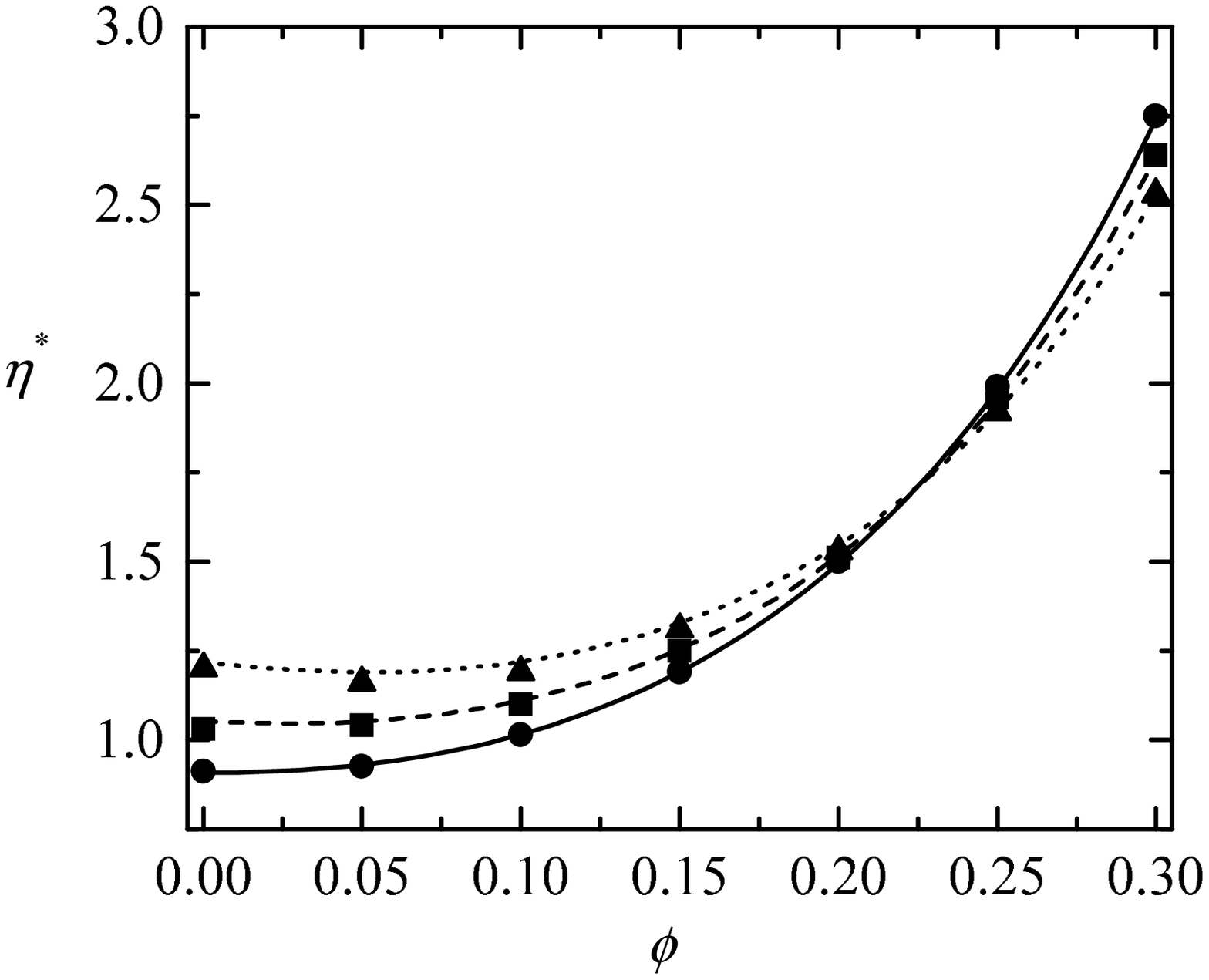}
\caption{Plot of the reduced shear viscosity $\eta^*$ as a function of the solid fraction $\phi$ for $m_1/m_2=4$, $\sigma_1/\sigma_2=1$, and $x_1=1/2$ and three different values of the restitution coefficient $\alpha$: $\alpha=0.9$ (solid line and circles), $\alpha=0.8$ (dashed line and squares), and $\alpha=0.7$ (dotted line and triangles). The lines are the theoretical predictions and the symbols refer to the results obtained from Monte Carlo simulations.
\label{fig5}}
\end{figure}

\begin{figure}
\includegraphics[width=0.5 \columnwidth]{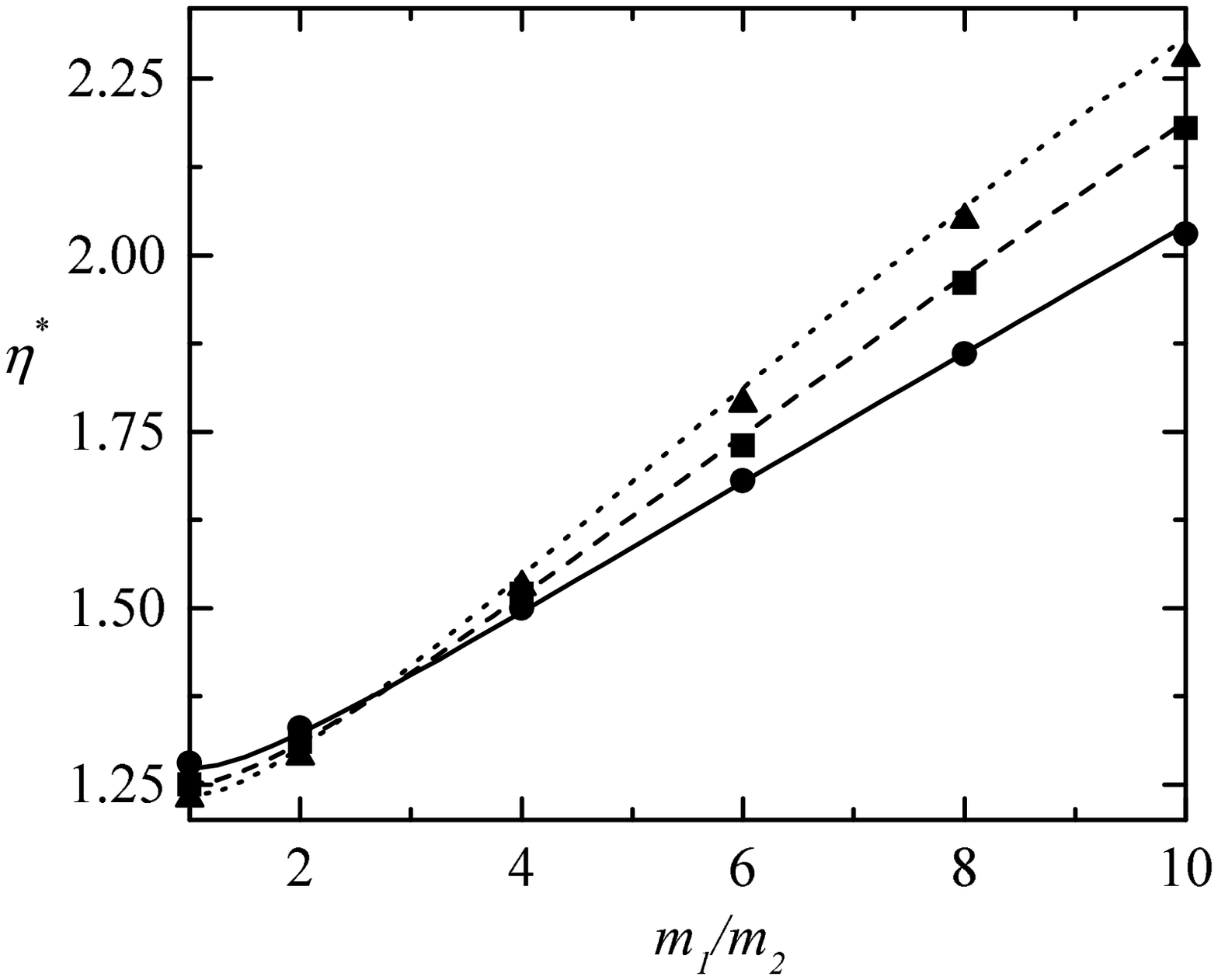}
\caption{Plot of the reduced shear viscosity $\eta^*$ as a function of the mass ratio $m_1/m_2=$, for  $\phi=0.2$, $\sigma_1/\sigma_2=1$, $x_1=1/2$ and three different values of the restitution coefficient $\alpha$: $\alpha=0.9$ (solid line and circles), $\alpha=0.8$ (dashed line and squares), and $\alpha=0.7$ (dotted line and triangles). The lines are the theoretical predictions and the symbols refer to the results obtained from Monte Carlo simulations.
\label{fig6}}
\end{figure}

\begin{figure}
\includegraphics[width=0.5 \columnwidth]{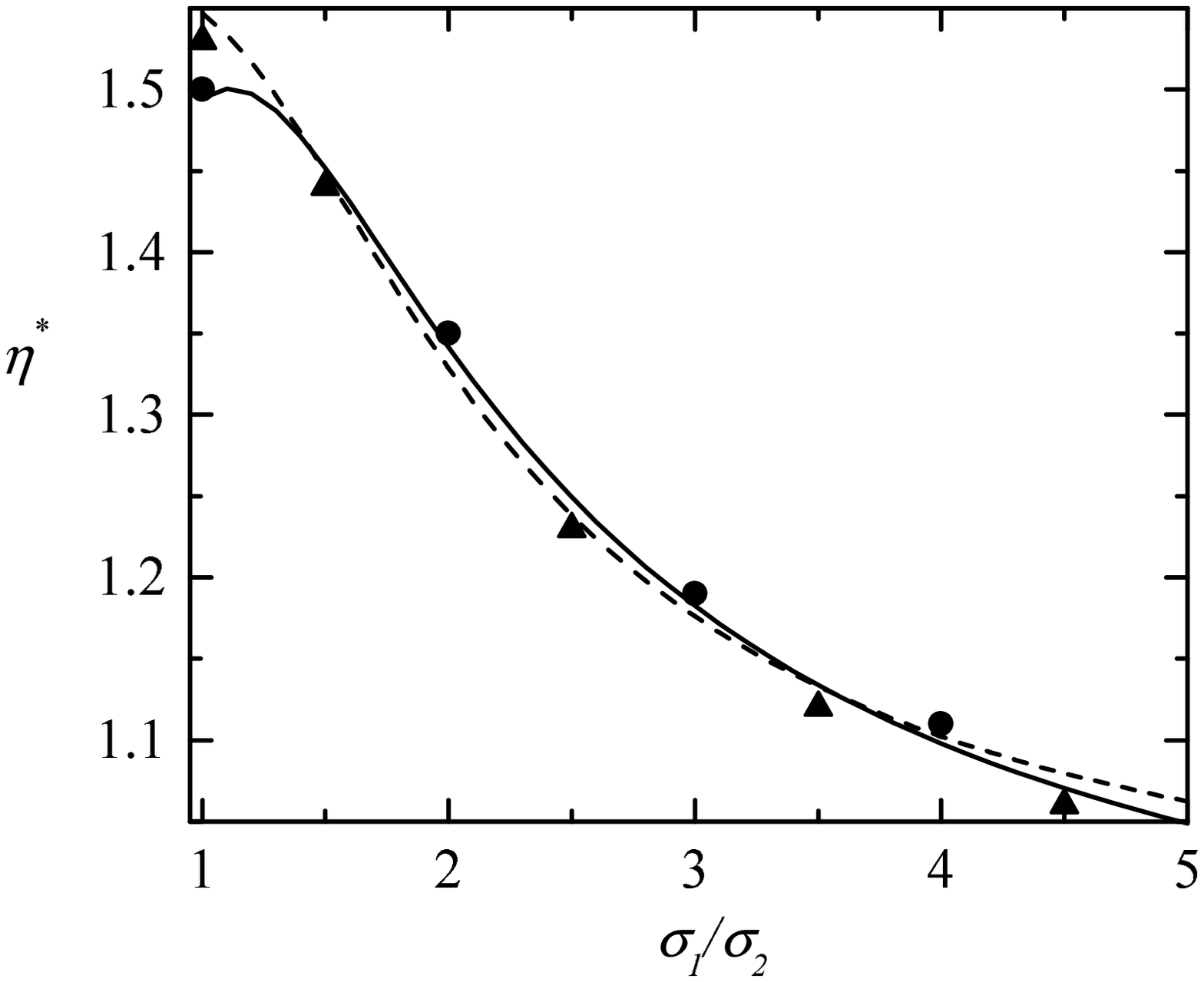}
\caption{Plot of the reduced shear viscosity $\eta^*$ as a function of the size ratio $\sigma_1/\sigma_2$ for $\phi=0.2$, $m_1/m_2=4$, $x_1=1/2$ and two different values of the restitution coefficient $\alpha$: $\alpha=0.9$ (solid line and circles) and $\alpha=0.7$ (dashed line and triangles). The lines are the theoretical predictions and the symbols refer to the results obtained from Monte Carlo simulations.
\label{fig7}}
\end{figure}

\begin{figure}
\includegraphics[width=0.5 \columnwidth]{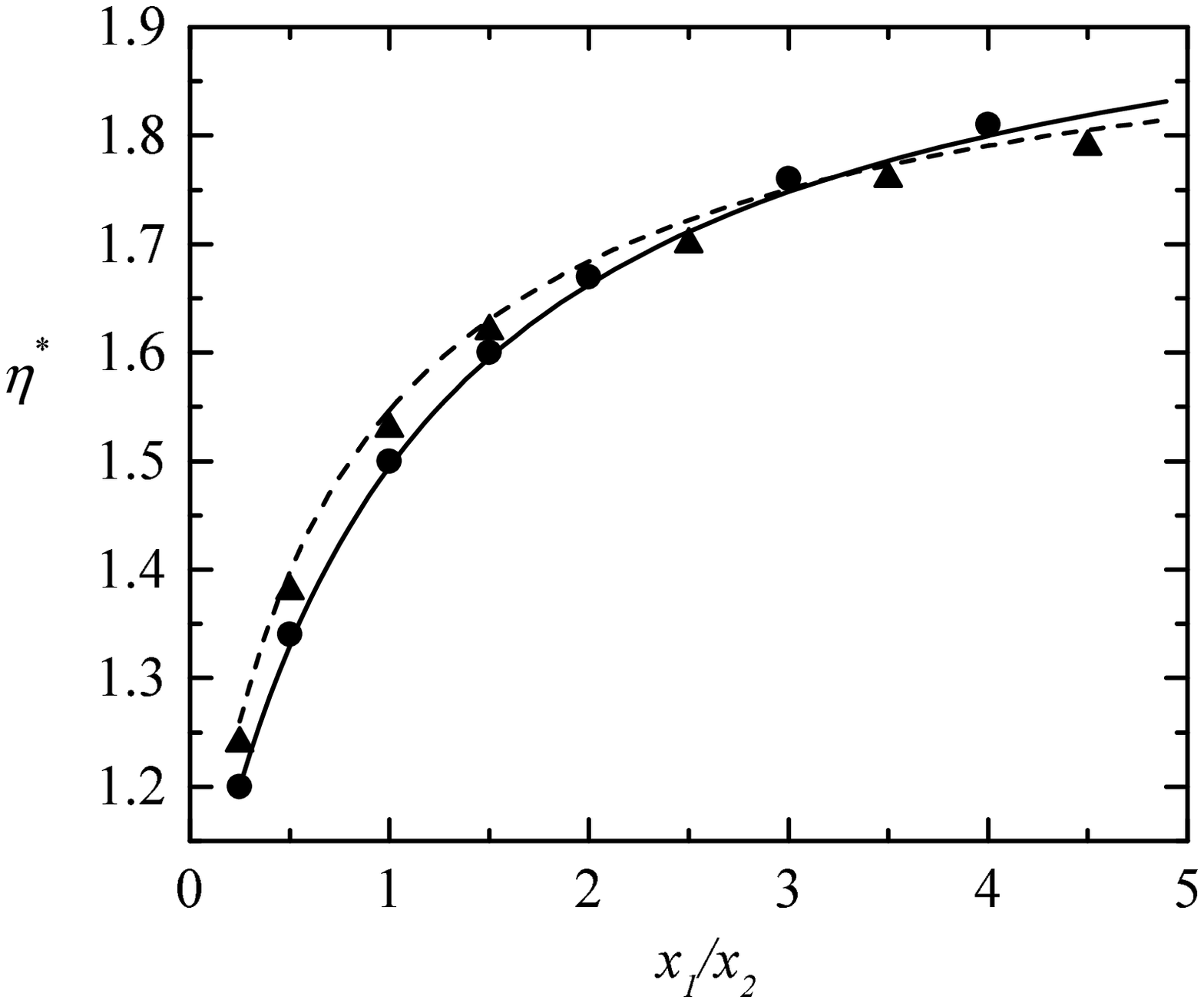}
\caption{Plot of the reduced shear viscosity $\eta^*$ as a function of the concentration ratio $x_1/x_2$ for $\phi=0.2$, $m_1/m_2=4$, $\sigma_1/\sigma_2=1$ and two different values of the restitution coefficient $\alpha$: $\alpha=0.9$ (solid line and circles) and $\alpha=0.7$ (dashed line and triangles). The lines are the theoretical predictions and the symbols refer to the results obtained from Monte Carlo simulations.
\label{fig8}}
\end{figure}

\subsection{Binary dense mixture}

Now, we consider granular binary mixtures whose particles can differ in size and mass.  First, to analyze density effects on the shear viscosity, in Figs.\ \ref{fig4} and \ref{fig5} the parameters of the mixture are $m_1/m_2=4$, $\sigma_1/\sigma_2=1$, and $x_1=1/2$. Three different values of $\alpha$ are studied: $\alpha=0.9$, 0.8, and 0.7.   The symbols are the same as in the previous figures. Figure \ref{fig4} shows the dependence of the kinetic part $\eta^{\text{k}*}=\nu \eta^{\text{k}}/nT$ on the solid fraction $\phi$, while the total shear viscosity $\eta^*$ is plotted in Fig.\ \ref{fig5}.  The good agreement between theory and simulation indicates that both kinetic and collisional transfer contributions are given accurately by the first Sonine approximation.  As in the monocomponent case [cf.\ Fig.\ \ref{fig2}], the shear viscosity of a granular mixture decreases (increases) as the inelasticity increases if the solid fraction is larger (smaller) than a given threshold value $\phi_0$. The value of $\phi_0$ depends on the parameters of the mixture although it is practically independent of dissipation. For the mixture considered in Fig.\ \ref{fig5}, $\phi_0(\alpha)\simeq 0.22$.

Next, we explore the influence of dissipation on the reduced shear viscosity $\eta^*$ for different  values of the mass ratio, the size ratio, and the mole fraction. We consider a solid fraction $\phi=0.2$ and different values of the restitution coefficient. In Fig.\ \ref{fig6} we plot $\eta^*$ versus the mass ratio $m_1/m_2$ for $\sigma_1/\sigma_1=1$ and $x_1=1/2$. As in the low-density case\cite{MG03}, we see that the influence of dissipation on $\eta^*$ becomes important  as the mass disparity increases. At a given value of the mass ratio, $\eta^*$ decreases (increases) with dissipation if the mass ratio is smaller (larger) than a certain threshold value, which value seems to be again practically independent of the restitution coefficient. Regarding the comparison between kinetic theory and simulation, we see that the agreement between both approaches is similar to the one previously obtained,  although the discrepancies tend to increase as $\alpha$ decreases. Figure \ref{fig7} shows the results for $\eta^*$ as a function of the size ratio for $m_1/m_2=4$ and $x_1=1/2$. We observe that the influence of $\alpha$ on $\eta^*$ is less significant as the one found before in Fig.\ \ref{fig6} for the mass ratio. Finally, in Fig.\ \ref{fig8}, $\eta^*$ is plotted as a function of the concentration ratio  $x_1/x_2$ for $m_1/m_2=4$ and $\sigma_1/\sigma_2=1$. As in Fig.\ \ref{fig7}, theory and simulation predict a weak influence of dissipation on the shear viscosity over the range of values of composition considered. It is worthwhile noting that the trends observed in Figs.\ \ref{fig6}--\ref{fig8} for finite density are similar to those previously reported in the low-density limit\cite{MG03}.

\section{Discussion}
\label{sec6}

The main goal of this paper has been to determine the shear viscosity $\eta$ of a binary mixture of smooth inelastic hard spheres described by the Enskog equation. To get the dependence of $\eta$ on the parameters of the mixture, the special state of uniform shear flow (USF) has been considered.  The USF is characterized by constant partial densities, uniform temperature, and by a linear profile of the $x$-component of the flow velocity along the $y$-direction. The (constant) shear rate $a$ is the relevant nonequilibrium parameter of the problem. Two complementary approaches have been used. First, a normal solution to the Enskog equation  is obtained through first order in $a$ by means of the Chapman-Enskog method. As in the elastic case, the shear viscosity coefficient $\eta$ is given in terms of the solution of a set of coupled linear integral equations, which are solved approximately by taking the leading terms in a Sonine polynomial expansion. The explicit form of $\eta$ is given by Eqs.\ (\ref{3.25})--(\ref{3.29}) as a function of the restitution coefficients, the temperature, the total solid fraction, and the masses, sizes, and concentrations of the constituents of the granular mixture. Second, the Enskog equation has been numerically solved in the USF by using an extension of the well-known Direct Simulation Monte Carlo Method (DSMC) \cite{B94} of the Boltzmann equation. The simulation has been performed by introducing an external (Gaussian) force which heats the system to compensate for the energy lost in collisions. Due to the action of this external driving force, the shearing work still heats the mixture and so the reduced shear rate $a^*(t)$ goes to zero for long times. As a consequence, the system reaches a regime described by linear hydrodynamics and the shear-viscosity coefficient can be measured in the simulations. 

The analysis made here extends previous results \cite{MG03} obtained by the authors for a binary mixture at low-density.  As in the latter case, the comparison between the Chapman-Enskog results in the first Sonine approximation and simulation data shows in general an excellent agreement for a wide range of values of densities, dissipation and parameters of the mixture. Discrepancies with simulation results are due mainly to the approximations carried out in the Chapman-Enskog scheme, and more specifically in taking only the first Sonine correction. However, apart from this source of slight discrepancy, the good agreement obtained here is a further testimony to the validity of a hydrodynamic description for granular media beyond the weak dissipation limit.  Moreover, a test of the utility of the Enskog theory at high densities is possible using molecular dynamics simulations. Previous comparisons at the level of partial temperatures\cite{DHGD02} and self-diffusion coefficient \cite{LBD02} indicate that the range of densities for which the RET applies decreases with increasing dissipation. We hope that the present results stimulate the performance of such simulations in the case of the shear viscosity.  

As in the low-density case, theory and simulation show that the dependence of $\eta$ on dissipation increases as the mass differences increase. The dependence of the shear viscosity on inelasticity is not significantly affected when composition and diameters are changed. With respect to the dependence on density, the results indicate that the shear viscosity of the granular fluid is larger than the one corresponding to a molecular fluid if the solid fraction $\phi$ is smaller than a threshold value $\phi_0$, while the opposite happens when $\phi>\phi_0$. The value of $\phi_0$ depends on the mechanical parameters of the mixture, but is practically independent of the restitution coefficients.  

Recently, a seemingly similar analysis on rheology of bidisperse granular mixtures has been carried out via event-driven simulations\cite{AL03}. However, this study is addressed to the steady sheared state achieved when viscous heating and collisional cooling exactly cancel each other. Under these conditions, due to the coupling between dissipation and the shear rate, the granular fluid is far away from the Navier-Stokes regime (non-Newtonian fluid), except when $\alpha\to 1$.  This precludes the possibility of making a comparison between the molecular dynamics results found in Ref.\ \cite{AL03} for the shear viscosity with the predictions of the Enskog equation. 

The main limitation of the results derived here is its restriction to the {\em uniform} shear flow state. The extension of this study to more {\em general} hydrodynamic states for a dense binary mixture, e.g. those with gradients of concentrations and temperature as well, is somewhat complex due to the technical difficulties of the Chapman-Enskog method associated with the spatial dependence of the pair correlation functions considered in the RET. We plan to extend the results derived years ago from the RET by L\'opez de Haro {\em et al.}\cite{MCK83} for a mixture of smooth elastic particles  to the case of {\em inelastic} collisions. Once the complete hydrodynamic equations of the mixture is known, some insight could be gained into the understanding of phenomena very often observed in nature and experiments, such as separation or segregation. 

\acknowledgments

V. G. acknowledges partial support from the Ministerio de Ciencia y Tecnolog\'{\i}a (Spain) thorugh Grant No. BFM2001-0718.

\appendix
\section{Collisional transfer contributions} 
\label{appA}

In this Appendix some details of the derivation of the collisional transfer contributions to the heat and momentum fluxes are given. First, we consider the collisional momentum transfer:
\begin{eqnarray}
\label{a1}
{\bf I}_p &\equiv& \sum_{i=1}^2\sum_{j=1}^2\int d{\bf v}_1 m_i {\bf v}_1 J_{ij}^{\text{E}}\left[ {\bf r}, {\bf v}_{1}|f_i,f_j\right] \nonumber\\
&=& \sum_{i=1}^2\sum_{j=1}^2\sigma_{ij}^{2}\int d{\bf v}_1\int d{\bf v} 
_{2}\int d\widehat{\boldsymbol {\sigma }}\,\Theta (\widehat{{\boldsymbol {\sigma }}} 
\cdot {\bf g})(\widehat{\boldsymbol {\sigma }}\cdot {\bf g}) m_i{\bf v}_1 \nonumber 
\\ &&\times \left[ \alpha_{ij} ^{-2}\chi_{ij}({\bf r},{\bf r}-\boldsymbol{\sigma }_{ij}) 
f_i({\bf r}, {\bf v}_1';t)f_j({\bf r}-\boldsymbol {\sigma}_{ij}, {\bf v}_2';t)\right.
\nonumber\\
& & \left.
-\chi_{ij}({\bf r},{\bf r}+\boldsymbol {\sigma}_{ij}) 
f_i({\bf r}, {\bf v}_1;t)f_j({\bf r}+\boldsymbol {\sigma}_{ij}, {\bf v}_2;t)\right].
\end{eqnarray} 
Now, we change variables to integrate over ${\bf v}_1'$ and ${\bf v}_2'$ instead of  ${\bf v}_1$ and ${\bf v}_2$ in the first term of the right-hand side  of (\ref{a1}). The Jacobian of the transformation is $\alpha_{ij}$ and $\widehat{\boldsymbol {\sigma }}\cdot {\bf g} =-\alpha_{ij}(\widehat{\boldsymbol {\sigma }}\cdot {\bf g}')$. Also, ${\bf v}_1({\bf v}_1', {\bf v}_2')\equiv {\bf v}_1''={\bf v}_1-\mu_{ji} (1+\alpha_{ij})\widehat{\boldsymbol {\sigma }}(\widehat{\boldsymbol {\sigma }}\cdot {\bf g})$ and in addition, we make the change $\widehat{\boldsymbol {\sigma }}\to -\widehat{\boldsymbol {\sigma }}$. Thus, the integral (\ref{a1}) becomes 
\begin{eqnarray}
\label{a2}
{\bf I}_p &=& \sum_{i=1}^2\sum_{j=1}^2\sigma_{ij}^{2}\int d{\bf v}_1\int d{\bf v} 
_{2}\int d\widehat{\boldsymbol {\sigma }}\,\Theta (\widehat{{\boldsymbol {\sigma }}} 
\cdot {\bf g})(\widehat{\boldsymbol {\sigma }}\cdot {\bf g}) m_i({\bf v}_1''-{\bf v}_1) \nonumber\\
& & \times \chi_{ij}({\bf r},{\bf r}+\mathbf{\sigma}_{ij}) 
f_i({\bf r}, {\bf v}_1;t)f_j({\bf r}+\boldsymbol {\sigma}_{ij}, {\bf v}_2;t).
\end{eqnarray} 
Since $m_i({\bf v}_1''-{\bf v}_1)=m_j({\bf v}_2-{\bf v}_2'')$, Eq.\ (\ref{a2}) can be rewritten as 
\begin{eqnarray}
\label{a3}
{\bf I}_p &=& \sum_{i=1}^2\sum_{j=1}^2\sigma_{ij}^{2}\int d{\bf v}_1\int d{\bf v} 
_{2}\int d\widehat{\boldsymbol {\sigma }}\,\Theta (\widehat{{\boldsymbol {\sigma }}} 
\cdot {\bf g})(\widehat{\boldsymbol {\sigma }}\cdot {\bf g}) m_j({\bf v}_2-{\bf v}_2'') \nonumber\\
& & \times \chi_{ij}({\bf r},{\bf r}+\mathbf{\sigma}_{ij}) 
f_i({\bf r}, {\bf v}_1;t)f_j({\bf r}+\boldsymbol {\sigma }_{ij}, {\bf v}_2;t) \nonumber\\
&=& \sum_{i=1}^2\sum_{j=1}^2\sigma_{ij}^{2}\int d{\bf v}_1\int d{\bf v} 
_{2}\int d\widehat{\boldsymbol {\sigma }}\,\Theta (\widehat{{\boldsymbol {\sigma }}} 
\cdot {\bf g})(\widehat{\boldsymbol {\sigma }}\cdot {\bf g}) m_i({\bf v}_1-{\bf v}_1'') 
\nonumber\\
& & \times
\chi_{ij}({\bf r}-\boldsymbol{\sigma }_{ij},{\bf r}) f_j({\bf r}, {\bf v}_2;t)f_i({\bf r}-\boldsymbol {\sigma }_{ij}, {\bf v}_1;t), 
\end{eqnarray} 
where in the last step we have exchanged the roles of the species $i$ and $j$, which implies the changes  $\widehat{\boldsymbol {\sigma }}\to -\widehat{\boldsymbol {\sigma }}$ and ${\bf v}_1\leftrightarrow {\bf v}_2$. Combination of Eqs.\ (\ref{a2}) and (\ref{a3}) yields 
\begin{eqnarray}
\label{a4}
{\bf I}_p &=&\sum_{i=1}^2\sum_{j=1}^2 \frac{\sigma_{ij}^{2}}{2}\int d{\bf v}_1\int d{\bf v} 
_{2}\int d\widehat{\boldsymbol {\sigma }}\,\Theta (\widehat{{\boldsymbol {\sigma }}} 
\cdot {\bf g})(\widehat{\boldsymbol {\sigma }}\cdot {\bf g}) m_i({\bf v}_1''-{\bf v}_1) \nonumber 
\\ &&\times \left[\chi_{ij}({\bf r},{\bf r}+\boldsymbol{\sigma }_{ij}) 
f_i({\bf r}, {\bf v}_1;t)f_j({\bf r}+\boldsymbol {\sigma}_{ij}, {\bf v}_2;t)\right. \nonumber\\
& & \left.
-\chi_{ij}({\bf r}-\boldsymbol{\sigma }_{ij},{\bf r})f_i({\bf r}-\boldsymbol {\sigma}_{ij}, {\bf v}_1;t)
 f_j({\bf r}, {\bf v}_2;t)\right].
\end{eqnarray}
Let $F_{ij}({\bf r}_1,{\bf r}_2)$ be the function
\begin{equation}
\label{a5}
F_{ij}({\bf r}_1,{\bf r}_2)=\chi_{ij}({\bf r}_1,{\bf r}_2)f_i({\bf r}_1)f_j({\bf r}_2).
\end{equation}
Next, we use the identity
\begin{eqnarray}
\label{a6}
F_{ij}({\bf r},{\bf r}+\boldsymbol {\sigma}_{ij})-F_{ij}({\bf r}-\boldsymbol {\sigma}_{ij},{\bf r})&=&
\int_{0}^{1}d\lambda \frac{\partial}{\partial \lambda}F_{ij}\left({\bf r}-(1-\lambda)\boldsymbol {\sigma}_{ij}, 
{\bf r}+\lambda \boldsymbol {\sigma}_{ij}\right)\nonumber\\
&=&\boldsymbol {\sigma}_{ij}\cdot \nabla \int_{0}^{1}d\lambda F_{ij}\left({\bf r}-(1-\lambda)\boldsymbol {\sigma}_{ij}, 
{\bf r}+\lambda \boldsymbol {\sigma}_{ij}\right).
\end{eqnarray}
Therefore, from Eqs.\ (\ref{a5}) and (\ref{a6}), Eq.\ (\ref{a4}) can be rewriten as a divergence:
\begin{eqnarray}
\label{a7}
{\bf I}_p &=& -\nabla\cdot \sum_{i=1}^2\sum_{j=1}^2\frac{\sigma_{ij}^{3}}{2}
\int d{\bf v}_1\int d{\bf v} _{2}\int d\widehat{\boldsymbol {\sigma }}\,\Theta (\widehat{{\boldsymbol {\sigma }}} 
\cdot {\bf g})(\widehat{\boldsymbol {\sigma }}\cdot {\bf g}) \widehat{\boldsymbol {\sigma}} \nonumber 
\\ &&\times m_i({\bf v}_1-{\bf v}_1'') \int_{0}^{1} d\lambda \chi_{ij}[{\bf r}-(1-\lambda)\boldsymbol {\sigma}_{ij},{\bf r}+\lambda \boldsymbol{\sigma }_{ij}] \nonumber\\
& & \times 
f_i({\bf r}- (1-\lambda)\boldsymbol {\sigma}_{ij},{\bf v}_1;t)f_j({\bf r}+\lambda \boldsymbol {\sigma}_{ij}, {\bf v}_2;t)\nonumber\\
&=&-\nabla\cdot \sum_{i=1}^2\sum_{j=1}^2 \sigma_{ij}^{3}\frac{m_im_j}{m_i+m_j}\frac{1+\alpha_{ij}}{2}
\int d{\bf v}_1\int d{\bf v} 
_{2}\int d\widehat{\boldsymbol {\sigma }}\,\Theta (\widehat{{\boldsymbol {\sigma }}} 
\cdot {\bf g})(\widehat{\boldsymbol {\sigma }}\cdot {\bf g})^2 \widehat{{\boldsymbol {\sigma }}} 
\widehat{{\boldsymbol {\sigma }}} \nonumber \\
&&\times  \int_{0}^{1} d\lambda \chi_{ij}[{\bf r}-(1-\lambda)\boldsymbol {\sigma}_{ij},{\bf r}+\lambda \mathbf{\sigma}_{ij}] f_i({\bf r}- (1-\lambda)\boldsymbol {\sigma}_{ij},{\bf v}_1;t)f_j({\bf r}+
\lambda \boldsymbol {\sigma}_{ij}, {\bf v}_2;t).\nonumber\\
\end{eqnarray}
According to the momentum balance equation (\ref{2.8}), the divergence of the collisional transfer part ${\sf P}^{\text{c}}$ is
\begin{equation}
\label{a8}
\sum_{i=1}^2\sum_{j=1}^2\int d{\bf v} m_i ({\bf v}-{\bf u})J_{ij}^{\text{E}}[{\bf v}|f_i,f_j]=-\nabla\cdot {\sf P}^{\text{c}}.
\end{equation}
By taking into account (\ref{a8}), one directly gets the expression (\ref{2.17}) for the collisional part of the pressure tensor:
\begin{eqnarray}
\label{a9}
{\sf P}^{\text{c}}({\bf r},t)&=&\sum_{i=1}^2\sum_{j=1}^2\sigma_{ij}^{3}\frac{m_im_j}{m_i+m_j}\frac{1+\alpha_{ij}}{2} 
\int d{\bf v}_1\int d{\bf v} _{2}\int d\widehat{\boldsymbol {\sigma }}\,\Theta (\widehat{{\boldsymbol {\sigma }}} 
\cdot {\bf g})(\widehat{\boldsymbol {\sigma }}\cdot {\bf g})^2 \widehat{{\boldsymbol {\sigma }}} \widehat{{\boldsymbol {\sigma }}}  \nonumber 
\\ &&\times  \int_{0}^{1} d\lambda \chi_{ij}[{\bf r}-(1-\lambda)\boldsymbol {\sigma}_{ij},{\bf r}+\lambda \mathbf{\sigma}_{ij}]f_i({\bf r}- (1-\lambda)\boldsymbol {\sigma}_{ij},{\bf v}_1;t)f_j({\bf r}+
\lambda \boldsymbol {\sigma}_{ij}, {\bf v}_2;t).\nonumber\\
\end{eqnarray}

The collisional transfer contribution to the energy balance equation is 
\begin{equation}
\label{a10}
{\bf I}_e \equiv \sum_{i=1}^2\sum_{j=1}^2\int d{\bf v}_1 \frac{m_i}{2} {\bf v}_1^2 J^{\text{E}}\left[ {\bf r}, {\bf v}_{1}|f_i,f_j\right]. 
\end{equation}
By following similar mathematical steps as in the case of momentum, one gets
\begin{eqnarray}
\label{a11}
{\bf I}_e&=& \sum_{i=1}^2\sum_{j=1}^2\frac{\sigma_{ij}^{2}}{2}\int d{\bf v}_1\int d{\bf v} 
_{2}\int d\widehat{\boldsymbol {\sigma }}\,\Theta (\widehat{{\boldsymbol {\sigma }}} 
\cdot {\bf g})(\widehat{\boldsymbol {\sigma }}\cdot {\bf g}) \frac{m_i}{2}({\bf v}_1^{''2}-{\bf v}_1) \nonumber 
\\ 
&&\times \left[\chi_{ij}({\bf r},{\bf r}+\boldsymbol{\sigma }_{ij}) 
f_i({\bf r}, {\bf v}_1;t)f_j({\bf r}+\boldsymbol {\sigma}_{ij}, {\bf v}_2;t)\right. \nonumber\\
& & \left.
-\chi_{ij}({\bf r},{\bf r}-\mathbf{\sigma}_{ij})f_i({\bf r}-\boldsymbol {\sigma }_{ij}, {\bf v}_1;t)
 f_j({\bf r}, {\bf v}_2;t)\right]\nonumber\\
& & 
-\sum_{i=1}^2\sum_{j=1}^2\sigma_{ij}^{2}\frac{m_im_j}{m_i+m_j}
\frac{(1-\alpha_{ij}^2)}{4}\int d{\bf v}_1\int d{\bf v} 
_{2}\int d\widehat{\boldsymbol {\sigma }}\,\Theta (\widehat{{\boldsymbol {\sigma }}} 
\cdot {\bf g})(\widehat{\boldsymbol {\sigma }}\cdot {\bf g})^3 \nonumber\\
& & \times
\chi_{ij}({\bf r},{\bf r}+\boldsymbol{\sigma }_{ij}) 
f_i({\bf r}, {\bf v}_1;t)f_j({\bf r}+\boldsymbol {\sigma}_{ij}, {\bf v}_2;t)\nonumber\\
&=&
-\nabla\cdot \frac{\sigma_{ij}^{3}}{2}\int d{\bf v}_1\int d{\bf v} 
_{2}\int d\widehat{\boldsymbol {\sigma }}\,\Theta (\widehat{{\boldsymbol {\sigma }}} 
\cdot {\bf g})(\widehat{\boldsymbol {\sigma }}\cdot {\bf g}) \widehat{{\boldsymbol {\sigma }}}  \nonumber \\
 &&\times \frac{m_i}{2}({\bf v}_1^2-{\bf v}_1^{''2}) \int_{0}^{1} d\lambda \chi_{ij}[{\bf r}-(1-\lambda)\boldsymbol {\sigma}_{ij},{\bf r}+\lambda \mathbf{\sigma}_{ij}) 
\nonumber\\
& & \times
f_i({\bf r}- (1-\lambda)\boldsymbol {\sigma}_{ij},{\bf v}_1;t)f_j({\bf r}+\lambda \boldsymbol {\sigma}_{ij}, {\bf v}_2;t)\nonumber\\
& & 
-\sum_{i=1}^2\sum_{j=1}^2\sigma_{ij}^{2}\frac{m_im_j}{m_i+m_j}
\frac{(1-\alpha_{ij}^2)}{4}\int d{\bf v}_1\int d{\bf v} 
_{2}\int d\widehat{\boldsymbol {\sigma }}\,\Theta (\widehat{{\boldsymbol {\sigma }}} 
\cdot {\bf g})(\widehat{\boldsymbol {\sigma }}\cdot {\bf g})^3 \nonumber\\ 
& & 
\times \chi_{ij}({\bf r},{\bf r}+\boldsymbol{\sigma }_{ij}) 
f_i({\bf r}, {\bf v}_1;t)f_j({\bf r}+\boldsymbol {\sigma}_{ij}, {\bf v}_2;t).
\end{eqnarray}
Here, we have used the identity (\ref{a6}) and the scattering law
\begin{equation}
\label{a12}
m_i({\bf v}_1^{''2}-{\bf v}_1^2)=m_j({\bf v}_2^2-{\bf v}_2^{''2})-\frac{m_im_j}{m_i+m_j}(1-\alpha_{ij}^2)  
(\widehat{\boldsymbol {\sigma }}\cdot {\bf g})^2.
\end{equation}
The balance energy equation (\ref{2.9}) yields 
\begin{equation}
\label{a13}
\sum_{i=1}^2\sum_{j=1}^2\int d{\bf v} \frac{m_i}{2} ({\bf v}-{\bf u})^2 J_{ij}^{\text{E}}[{\bf v}|f_i,f_j]=-\nabla \cdot {\bf q}^{\text{c}}-{\sf P}^{\text{c}}:\nabla {\bf u}-\frac{3}{2}nT \zeta, 
\end{equation}
where ${\bf q}^{\text{c}}$ is the collisional contribution to the heat flux and $\zeta$ is the cooling rate. To identify such quantities, we use the relation 
\begin{equation}
\label{a14}
\frac{m_i}{2}\left({\bf v}_1^2-{\bf v}_1^{''2}\right)=\frac{m_i}{2}\mu_{ji}^2(1-\alpha_{ij}^2)(\widehat{\boldsymbol {\sigma }}\cdot {\bf g})^2+m_i\mu_{ji}(1+\alpha_{ij})
(\widehat{\boldsymbol {\sigma }}\cdot {\bf g})[(\widehat{\boldsymbol {\sigma }}\cdot {\bf G}_{ij})+
(\widehat{\boldsymbol {\sigma }}\cdot {\bf u})],
\end{equation}
with ${\bf G}_{ij}=\mu_{ij}{\bf V}_1+\mu_{ji}{\bf V}_2$, ${\bf V}={\bf v}-{\bf u}$ being the peculiar velocity. Comparing Eqs.\ (\ref{a11}) and (\ref{a13}) and taking into account Eqs.\ (\ref{a7}) and (\ref{a14}), one can finally obtain the expressions (\ref{2.18}) and (\ref{2.19}) for ${\bf q}^{\text{c}}$ and $\zeta$. They are given by 
\begin{eqnarray}
\label{a15}
{\bf q}^{\text{c}}({\bf r},t)&=&\sum_{i=1}^2\sum_{j=1}^2\sigma_{ij}^{3}\frac{m_im_j}{m_i+m_j}\frac{1+\alpha_{ij}}{2} 
\int d{\bf v}_1\int d{\bf v} _{2}\int d\widehat{\boldsymbol {\sigma }}\,\Theta (\widehat{{\boldsymbol {\sigma }}} 
\cdot {\bf g})(\widehat{\boldsymbol {\sigma }}\cdot {\bf g})^2 \widehat{{\boldsymbol {\sigma }}} \nonumber\\ 
& & \times
\left[(\widehat{\boldsymbol {\sigma }}\cdot {\bf G}_{ij})+\frac{1}{2}\mu_{ji}(1-\alpha_{ij})
(\widehat{\boldsymbol {\sigma }}\cdot {\bf g})\right]\nonumber\\ 
&&\times  \int_{0}^{1} d\lambda \chi_{ij}[{\bf r}-(1-\lambda)\boldsymbol {\sigma}_{ij},{\bf r}+\lambda \mathbf{\sigma}_{ij}) \nonumber\\
& & \times f_i({\bf r}- (1-\lambda)\boldsymbol {\sigma}_{ij},{\bf v}_1;t)f_j({\bf r}+\lambda \boldsymbol {\sigma}_{ij}, {\bf v}_2;t) \nonumber\\
&=&\sum_{i=1}^2\sum_{j=1}^2\sigma_{ij}^{3}\frac{m_im_j}{m_i+m_j}\frac{1+\alpha_{ij}}{2} 
\int d{\bf v}_1\int d{\bf v} 
_{2}\int d\widehat{\boldsymbol {\sigma }}\,\Theta (\widehat{{\boldsymbol {\sigma }}} 
\cdot {\bf g})(\widehat{\boldsymbol {\sigma }}\cdot {\bf g})^2 \widehat{{\boldsymbol {\sigma }}}\nonumber\\
& & \times   
\left[(\widehat{\boldsymbol {\sigma }}\cdot {\bf G}_{ij})+\frac{1}{4}\left(\mu_{ji}-\mu_{ij}\right)(1-\alpha_{ij})
(\widehat{\boldsymbol {\sigma }}\cdot {\bf g})\right]\nonumber\\ 
&&\times  \int_{0}^{1} d\lambda \chi_{ij}[{\bf r}-(1-\lambda)\boldsymbol {\sigma}_{ij},{\bf r}+\lambda \mathbf{\sigma}_{ij}) f_i({\bf r}- (1-\lambda)\boldsymbol {\sigma}_{ij},{\bf v}_1;t)f_j({\bf r}+\lambda \boldsymbol {\sigma}_{ij}, {\bf v}_2;t),\nonumber\\
\end{eqnarray}
\begin{eqnarray}
\label{a16}
\zeta({\bf r},t)&=&\frac{1}{6nT}\sum_{i=1}^2\sum_{j=1}^2\sigma_{ij}^{2}\frac{m_im_j}{m_i+m_j}
(1-\alpha_{ij}^2)
\int d{\bf v}_1\int d{\bf v} _{2}\int d\widehat{\boldsymbol {\sigma }}\,\Theta (\widehat{{\boldsymbol {\sigma }}} 
\cdot {\bf g})(\widehat{\boldsymbol {\sigma }}\cdot {\bf g})^3 \nonumber\\
& & \times
\chi_{ij}({\bf r},{\bf r}+\boldsymbol{\sigma}_{ij}) 
f_i({\bf r}, {\bf v}_1;t)f_j({\bf r}+\boldsymbol {\sigma}_{ij}, {\bf v}_2;t).
\end{eqnarray}
The second equality in Eq.\ (\ref{a15}) has been obtained by exchanging the roles of species $i$ and $j$. In the case of mechanically equivalent particles, Eqs.\ (\ref{a9}), (\ref{a15}), and (\ref{a16}) reduce to those previously obtained  for a monocomponent dense gas\cite{BDS97}.

\section{First order solution to the USF}
\label{appB}

In this Appendix we apply the Chapman-Enskog method to solve Eq.\ (\ref{3.4}) to first order in the shear rate $a$. First, in order to get the kinetic equation for $f_1^{(1)}$,  the Enskog collision operator (\ref{3.5}) must be expanded as  
\begin{eqnarray}
\label{b1}
J_{ij}^{\text{E}}&\to& J_{ij}^{(0)}[f_i^{(0)},f_j^{(0)}]+J_{ij}^{(0)}[f_i^{(1)},f_j^{(0)}]+J_{ij}^{(0)}[f_i^{(0)},f_j^{(1)}]
\nonumber\\
& & +a \Lambda_{ij}[f_i^{(0)},f_j^{(0)}],
\end{eqnarray}
where the (Boltzmann) collision operator $J_{ij}^{(0)}[X_1,X_2]$ is defined in Eq.\ (\ref{3.12}) and 
\begin{eqnarray}
\label{b2}
\Lambda_{ij}[{\bf V}_1|f_i^{(0)},f_j^{(0)}]&=&\chi_{ij}\sigma_{ij}^{3} \int d{\bf V} 
_{2}\int d\widehat{\boldsymbol {\sigma }}\,\Theta (\widehat{{\boldsymbol {\sigma }}} 
\cdot {\bf g})(\widehat{\boldsymbol {\sigma }}\cdot {\bf g}) \widehat{\sigma }_y\left[
\alpha_{ij}^{-2}f_i^{(0)}({\bf V}_1')\frac{\partial}{\partial V_{2x}'}f_j^{(0)}({\bf V}_2')\right.
\nonumber\\
& & \left. + f_i^{(0)}({\bf V}_1)\frac{\partial}{\partial V_{2x}}f_j^{(0)}({\bf V}_2)\right].
\end{eqnarray}

Therefore, the distribution $f_i^{(1)}$ verifies the equation 
\begin{equation} 
\label{b3}
\left(\partial _{t}^{(0)}+ \frac{1}{2}\xi^{(0)}
\frac{\partial}{\partial {\bf V}}\cdot {\bf V}+{\cal L}_i\right)f_i^{(1)}+{\cal M}_if_j^{(1)}
=aV_y\frac{\partial}{\partial V_x} 
f_{i}^{(0)}+a\sum_{j=1}^2\Lambda_{ij}[f_i^{(0)},f_j^{(0)}],
\end{equation}
where it is understood that $i\neq j$ on the left hand side and the linear operators ${\cal L}_i$ and ${\cal M}_i$ are 
\begin{equation}
\label{b4}
{\cal L}_if_i^{(1)}=-\left(J_{ii}[f_i^{(0)},f_i^{(1)}]+
J_{ii}[f_i^{(1)},f_i^{(0)}]+J_{ij}[f_i^{(1)},f_j^{(0)}]\right), 
\end{equation}
\begin{equation}
\label{b5}
{\cal M}_if_j^{(1)}=-J_{ij}[f_i^{(0)},f_j^{(1)}]. 
\end{equation}
In these equations, use has been made of the fact that $\partial_t^{(1)}f_i^{(0)}=0$ according to the second identity of Eq.\ (\ref{3.10bis}). Furthermore, $\zeta^{(1)}=0$ by symmetry because $\nabla \cdot {\bf u}=0$ in the USF. The action of the time derivative $\partial_t^{(0)}$ on $f_i^{(1)}$ can be easily obtained from (\ref{3.10bis}) as  
\begin{equation}
\label{b6}
\partial_t^{(0)}f_i^{(1)}=-(\zeta^{(0)}-\xi^{(0)})T\partial_Tf_i^{(1)},
\end{equation}
and so, the integral equation (\ref{b3}) finally becomes 
\begin{equation} 
\label{b7}
\left[-(\zeta^{(0)}-\xi^{(0)})T\partial_T+ \frac{1}{2}\xi^{(0)}
\frac{\partial}{\partial {\bf V}}\cdot {\bf V}+{\cal L}_i\right]f_i^{(1)}+{\cal M}_if_j^{(1)}=V_y\frac{\partial}{\partial V_x} f_{i}^{(0)}+\sum_{j=1}^2\Lambda_{ij}[f_i^{(0)},f_j^{(0)}].
\end{equation}
This is the result (\ref{3.19}) used in the text. 

\section{Evaluation of some collision integrals}
\label{appC}

In this Appendix some collision integrals appearing along the text are evaluated. First, let us consider the integral (\ref{3.28}) 
\begin{eqnarray}
\label{c1}
\widetilde{\Lambda}_{ij}&=&\frac{1}{n_iT_i^2}\int d{\bf V}_1 m_iV_{1x}V_{1y}\Lambda_{ij}[{\bf V}_1|f_i^{(0)},f_j^{(0)}]\nonumber\\
&=&\frac{m_i}{n_iT_i^2}\chi_{ij}\sigma_{ij}^{3}\int d{\bf V}_1\int d{\bf V} 
_{2}\int d\widehat{\boldsymbol {\sigma }}\,\Theta (\widehat{{\boldsymbol {\sigma }}} 
\cdot {\bf g})(\widehat{\boldsymbol {\sigma }}\cdot {\bf g}) \widehat{\sigma }_yV_{1x}V_{1y}\left[
\alpha_{ij}^{-2}f_i^{(0)}({\bf V}_1')\frac{\partial}{\partial V_{2x}'}f_j^{(0)}({\bf V}_2')\right.
\nonumber\\
& & \left. + f_i^{(0)}({\bf V}_1)\frac{\partial}{\partial V_{2x}}f_j^{(0)}({\bf V}_2)\right].
\end{eqnarray}
As done in Appendix \ref{appA}, we change variables to integrate over ${\bf V}_1'$ and ${\bf V}_2'$ instead of ${\bf V}_1$ and ${\bf V}_2$ in the first term of the right-hand side of (\ref{c1}). Thus, the integral becomes   
\begin{eqnarray}
\label{c2}
\widetilde{\Lambda}_{ij}&=&-\frac{m_i}{n_iT_i^2}\chi_{ij}\sigma_{ij}^{3}\int d{\bf V}_1\int d{\bf V}_{2}\int d\widehat{\boldsymbol {\sigma }}\,\Theta (\widehat{{\boldsymbol {\sigma }}} 
\cdot {\bf g})(\widehat{\boldsymbol {\sigma }}\cdot {\bf g}) \widehat{\sigma }_y
f_i^{(0)}({\bf V}_1)\frac{\partial}{\partial V_{2x}}f_j^{(0)}({\bf V}_2)
\nonumber\\
& & \times 
\left(V_{1x}''V_{1y}''-V_{1x}V_{1y}\right), 
\end{eqnarray}
where 
\begin{equation}
\label{c3}
{\bf V}_1''={\bf V}_1-\mu_{ji} (1+\alpha_{ij})\widehat{\boldsymbol {\sigma }}
(\widehat{\boldsymbol {\sigma }}\cdot {\bf g}).
\end{equation}
Using (\ref{c3}), the last term on the right hand side of (\ref{c2}) can be explicitly computed as 
\begin{eqnarray}
\label{c4}
V_{1x}''V_{1y}''-V_{1x}V_{1y}&=&-\mu_{ji}(1+\alpha_{ij})(\widehat{\boldsymbol {\sigma }}\cdot {\bf g}) 
\left[G_{ij,x}\widehat{\sigma }_y+G_{ij,y}\widehat{\sigma}_x\right. \nonumber\\
& & +\left. \mu_{ji}(g_x\widehat{\sigma}_y+g_y\widehat{\sigma}_x)-
\mu_{ji}(1+\alpha_{ij})(\widehat{\boldsymbol {\sigma }}\cdot {\bf g}) \widehat{\sigma}_x
\widehat{\sigma}_y\right],
\end{eqnarray}
where $G_{ij,x}=\mu_{ij}V_{1x}+\mu_{ji}V_{2x}$ and $G_{ij,y}=\mu_{ij}V_{1y}+\mu_{ji}V_{2y}$.  
Substitution of Eq.\ (\ref{c4}) into Eq.\ (\ref{c2}) allows the angular integral to be performed with the result
\begin{eqnarray}
\label{c5}
\int d\widehat{\boldsymbol {\sigma }}\,\Theta (\widehat{{\boldsymbol {\sigma }}} 
\cdot {\bf g})(\widehat{\boldsymbol {\sigma }}\cdot {\bf g}) \widehat{\sigma }_y
& & \left(V_{1x}''V_{1y}''-V_{1x}V_{1y}\right)=-\frac{2\pi}{15}\mu_{ji}(1+\alpha_{ij})\left[(2g_y^2+g^2)G_{ij,x}\right.
\nonumber\\
& & \left. +2g_xg_yG_{ij,y}+\frac{2}{7}\mu_{ji}(11-3\alpha_{ij})g_xg_y^2+\frac{1}{7}
\mu_{ji}(4-3\alpha_{ij})g_xg^2\right].
\end{eqnarray}
Using (\ref{c5}), the integral (\ref{c2}) becomes
\begin{eqnarray}
\label{c6}
\widetilde{\Lambda}_{ij}&=& -\frac{2\pi}{15}\frac{m_i}{n_iT_i^2}\chi_{ij}\sigma_{ij}^{3}\mu_{ji}(1+\alpha_{ij})
\int d{\bf V}_1\int d{\bf V}_2f_i^{(0)}({\bf V}_1) f_j^{(0)}({\bf V}_2) \nonumber\\
& & \times\left[\frac{\mu_{ji}}{3}(3\alpha_{ij}-1)(V_1^2+V_2^2)-\frac{4}{3}\left(\mu_{ij}V_1^2-
\mu_{ji}V_2^2\right)\right] \nonumber\\
&=&-\frac{2\pi}{15}\frac{m_in_j}{T_i^2}\chi_{ij}\sigma_{ij}^{3}\mu_{ji}(1+\alpha_{ij})\left[\mu_{ji}
(3\alpha_{ij}-1)\left(\frac{T_i}{m_i}+\frac{T_j}{m_j}\right)-4\frac{T_i-T_j}{m_i+m_j}\right].
\end{eqnarray}
In the case of mechanically equivalent particles,  Eq.\ (\ref{c6}) coincides with the one previously obtained in the context of determining the shear viscosity in a monocomponent granular gas \cite{GD99a}.  

The collision frequencies $\tau_{ij}$ defined by the integrals (\ref{3.26}) and (\ref{3.27}) are the same as those appearing in the Boltzmann limit (except for the factors $\chi_{ij}$)\cite{GD02,MG03}. The details will not be repeated here and only the results are displayed. They are given by   
\begin{eqnarray}
\label{c7}
\tau_{11}&=&\frac{16}{5}\sqrt{\frac{\pi T_1}{m_1}}n_1\sigma_{1}^2
\chi_{11}\left[1-\frac{1}{4}(1-\alpha_{11})^2\right]
\left(1-\frac{c_1}{64}
\right)\nonumber\\
& & +\frac{8}{15}\sqrt{\pi}n_2\sigma_{12}^2\mu_{21}\chi_{12}v_0
(1+\alpha_{12})\theta_1^{3/2}\theta_2^{-1/2}
\left[6\theta_1^{-2}(\mu_{12}\theta_2-\mu_{21}\theta_1)
(\theta_1+\theta_2)^{-1/2}\right.\nonumber\\
& & +\frac{3}{2}\mu_{21}\theta_1^{-2}(\theta_1+\theta_2)^{1/2}(3-\alpha_{12})
+5\theta_1^{-1}(\theta_1+\theta_2)^{-1/2}\nonumber\\
& & \left. +\frac{c_2}{16}\frac{2\theta_2(12\mu_{21}+9\mu_{12}-10)-
\theta_1(5-6\mu_{21})-\case{3}{2}\mu_{21}(3-\alpha_{12})(\theta_1+\theta_2)}
{(\theta_1+\theta_2)^{5/2}}\right],
\end{eqnarray}
\begin{eqnarray}
\label{c8}
\tau_{12}&=&\frac{8}{15}\sqrt{\pi}n_2\sigma_{12}^2\frac{\mu_{21}^2}{\mu_{12}}\chi_{12}v_0
(1+\alpha_{12})\theta_1^{3/2}\theta_2^{-1/2}
\left[6\theta_2^{-2}(\mu_{12}\theta_2-\mu_{21}\theta_1)
(\theta_1+\theta_2)^{-1/2}\right.\nonumber\\
& & +\frac{3}{2}\mu_{21}\theta_2^{-2}(\theta_1+\theta_2)^{1/2}(3-\alpha_{12})
-5\theta_2^{-1}(\theta_1+\theta_2)^{-1/2}\nonumber\\
& & \left. +\frac{c_1}{16}\frac{2\theta_1(10-12\mu_{12}-9\mu_{21})+
\theta_2(5-6\mu_{12})-\case{3}{2}\mu_{21}(3-\alpha_{12})(\theta_1+\theta_2)}
{(\theta_1+\theta_2)^{5/2}}\right].
\end{eqnarray}
The corresponding expressions for $\tau_{22}$ and $\tau_{21}$ can be 
inferred from Eqs.\ (\ref{c7}) and (\ref{c8}) by exchanging 
$1\leftrightarrow2$. 

Finally, the collision contribution to the shear viscosity is given by Eq.\ (\ref{3.23}). To get explicit results, we have to evaluate integrals of the form
\begin{equation}
\label{c9}
A_{ij}=\int d{\bf V}_1\int d{\bf V}_2f_i^{(0)}({\bf V}_1)f_j^{(0)}({\bf V}_2)g, 
\end{equation}
by using the leading Sonine approximation (\ref{3.17}). Let us consider the integral $A_{12}$. Substitution of Eq.\ (\ref{3.17}) into Eq.\ (\ref{c9})  and neglecting nonlinear terms in $c_i$, $A_{12}$ can be written as 
\begin{eqnarray}
\label{c10}
A_{12}&=&n_1n_2\pi^{-3}v_0 \left(\theta_1\theta_2\right)^{3/2}\left\{I(\theta_1,\theta_2)\right.
\nonumber\\
& & +\frac{c_1}{4}\left(\theta_1^2\frac{d^2}{d\theta_1^2}+5\theta_1\frac{d}{d\theta_1}+\frac{15}{4}\right)
I(\theta_1,\theta_2)\nonumber\\
& & \left. 
+\frac{c_2}{4}\left(\theta_2^2\frac{d^2}{d\theta_2^2}+5\theta_2\frac{d}{d\theta_2}+\frac{15}{4}\right)
I(\theta_1,\theta_2)\right\},
\end{eqnarray}
where the dimensionles integral $I(\theta_1,\theta_2)$ is 
\begin{equation}
\label{c11}
I(\theta_1,\theta_2)=\int d{\bf V}_1^*\int d{\bf V}_2^* e^{-\theta_1V_1^{*2}-\theta_2V_2^{*2}},
\end{equation}
with ${\bf V}^*={\bf V}/v_0$.  The integral $I(\theta_1,\theta_2)$ can be performed by the change of variables 
{\begin{equation}
\label{c12}
{\bf x}={\bf V}_1^*-{\bf V}_2^*, \quad {\bf y}=\theta_1{\bf V}_1^*+\theta_2{\bf V}_2^*,
\end{equation}
with the Jacobian $(\theta_1+\theta_2)^{-3}$. The integral becomes
\begin{equation}
\label{c13}
I(\theta_1,\theta_2)=2\pi^{5/2}\frac{(\theta_1+\theta_2)^{1/2}}{\theta_1^2\theta_2^2}.
\end{equation}
Use of this result in (\ref{c10}) gives
\begin{equation}
\label{c14}
A_{12}=A_{21}=\frac{2}{\sqrt{\pi}}n_1n_2v_0\left(\frac{\theta_1+\theta_2}{\theta_1\theta_2}\right)^{1/2}
\left[1-\frac{c_1}{16}\left(\frac{\theta_2}{\theta_1+\theta_2}\right)^{2}
-\frac{c_2}{16}\left(\frac{\theta_1}{\theta_1+\theta_2}\right)^{2}\right].
\end{equation}
The corresponding expressions for $A_{11}$ and $A_{22}$ can be easily inferred from Eq.\ (\ref{c14}) by exchanging $1\leftrightarrow 2$:
\begin{equation}
\label{c15}
A_{11}=\frac{4}{\sqrt{\pi}}n_1^2\sqrt{\frac{T_1}{m_1}}\left(1-\frac{c_1}{32}\right),
\end{equation}
\begin{equation}
\label{c16}
A_{22}=\frac{4}{\sqrt{\pi}}n_2^2\sqrt{\frac{T_2}{m_2}}\left(1-\frac{c_2}{32}\right).
\end{equation}
Equations (\ref{c14})--(\ref{c16}) lead directly to the result given by Eq.\ (\ref{3.29}) in the text.

\end{document}